\documentclass[draft]{agujournal2019}
\usepackage{apacite}
\usepackage{aas_macros}
\usepackage{amssymb}
\usepackage{url} 
\usepackage{lineno}
\usepackage{soul}

\draftfalse

\journalname{Geophysical Research Letters}

\begin{document}

%
%


\title{Modeling a Transient Secondary Paleo-Lunar Atmosphere: 3-D Simulations and Analysis}

%
%




\authors{I.~Aleinov\affil{1,2}, M.J.~Way\affil{2,3,4}, C.~Harman\affil{5,2}, K.~Tsigaridis\affil{1,2}, E.T.~Wolf\affil{6},
G.~Gronoff\affil{7,8}}

\affiliation{1}{Center for Climate Systems Research, Columbia University, New York, NY 10025, USA}
\affiliation{2}{NASA Goddard Institute for Space Studies, 2880 Broadway, New York, New York, 10025, USA}
\affiliation{3}{Theoretical Astrophysics, Department of Physics and Astronomy, Uppsala University, Uppsala, SE-75120, Sweden}
\affiliation{4}{GSFC Sellers Exoplanet Environments Collaboration}
\affiliation{5}{Department of Applied Physics and Applied Mathematics, Columbia University, New York, NY 10025, USA}
\affiliation{6}{University of Colorado, Boulder, USA}
\affiliation{7}{Science Directorate, Chemistry and Dynamics Branch, NASA Langley Research Center, Hampton, VA, USA}
\affiliation{8}{SSAI, Hampton, VA, USA}





\correspondingauthor{Igor Aleinov}{Igor.Aleinov@nasa.gov}




\begin{keypoints}
\item We confirm the viability of transient maria outgassed atmosphere for pressures from 1 mb to 10 mb
\item 3-D simulations demonstrate where volatile deposition may occur
\item Atmospheric escape \& outgassing chemistry play important roles
\end{keypoints}

\begin{abstract} \newline

The lunar history of water deposition, loss, and transport post-accretion has become an important consideration in relation to the possibility of a human outpost on the Moon. Very recent work has shown that a secondary primordial atmosphere of up to 10~mb could have been emplaced $\sim$3.5$\times$10$^{9}$ years ago due to volcanic outgassing from the maria. Using a zero dimensional chemistry model we demonstrate the temperature dependence of the resulting major atmospheric components (CO or CO$_2$). We use a three dimensional general circulation model to test the viability of such an atmosphere and derive its climatological characteristics. Based on these results we then conjecture on its capability to transport volatiles outgassed from the maria to the permanently shadowed regions at the poles. Our preliminary results demonstrate that atmospheres as low as 1~mb are viable and that permanent cold trapping of volatiles is only possible at the poles.

\end{abstract}

\section{Introduction}

The Moon has had a number of relatively short-lived atmospheres.
Soon after its formation it is likely to have had a 1000--2500~K lunar magma ocean (LMO). The initial primordial lunar atmosphere was formed as a result of outgassing from the LMO. Work by \citeA{Stern1999} claimed this SiO$_2$-dominated atmosphere could have been as thick as present day Venus' atmosphere with a scale height of order 75 km. More recent work by \citeA{Saxena2017} state that it was a metal-dominated atmosphere (with SiO$_2$ vapor being present for only $\sim$100 years after the formation) of 1--100 mb on the warm side and collapsing on the cold far side of the planet. Such an atmosphere would have been short-lived and collapsed completely once the LMO had a crystallized crust. According to \citeA{Saxena2017} the crystallization would have happened in $\sim$1000 years, while complete solidification of the LMO could have taken up to 10--100 million years (My) \cite{Nemchin2009,Elkins2011}. \citeA{Stern1999} postulated a second epoch in which mare volcanism gave rise to a thin atmosphere punctuated by a `high-mass' collisionally--supported atmosphere of order 10$^{-5}$ mb for very short periods of time. Other research \cite{Stewart2011,Prem2015} also suggests that a thin transient atmosphere could have been formed by volatile-rich impacts. 
For the rest of its history the Moon's atmosphere was an exceedingly thin exosphere, which currently measures $\sim$$10^{-12}$ mb \cite{Stern1999,Cook2013,Benna2015}.

More recent research by \citeA{Needham2017} (hereafter NK2017) explores in detail the second epoch mentioned above. The longevity of this short-term `high-mass' atmosphere would be controlled by a competition between outgassing from the maria and atmospheric escape. According to NK2017 most of the outgassing from lunar maria occurred between 3.8$\times$10$^{9}$ and 3.1$\times$10$^{9}$ years ago (Gya) and at its peak it could sustain an atmosphere of $\sim$10 mb for $\sim$70 million years. Such an atmosphere would be similar in thickness to that of modern Mars and could play a major role in transporting volatiles from the maria to the poles. In turn these volatiles could be trapped in permanently shadowed regions (PSR) or buried under the regolithic dust.

We have expanded the study of NK2017 by first looking at results from a zero-dimensional (0-D) chemistry model to determine the atmosphere's major constituents based on the species outgassed from the maria. We then use these constituents as input to a three dimensional (3-D) general circulation model (GCM) \cite{Way2017} to better understand the temperature structure of this hypothetical atmosphere. We hope to eventually use such 3-D simulations to better understand volatile transport, deposition, and ultimately, volatile distributions on the Moon.

In Section \ref{chemistry} we discuss the importance of chemistry in determining the possible composition of the transient atmosphere. In Section \ref{escape} we explore some of the nuances of atmospheric escape. Section \ref{model} outlines how the 3-D GCM was set up and Section \ref{discussion} describes the results and some of the broader implications.

\section{Chemical composition of the lunar atmosphere}\label{chemistry}

As described by NK2017 the volcanic gases entering the atmosphere from the maria were predominantly CO, H$_{2}$O, sulfur (S), and H$_{2}$. However, much of the water vapor would be subject to condensation soon after eruption, as determined by the surface temperature. Atomic S is not a conventionally volatile species \cite<e.g.,>{Kasting1989}, and other studies have suggested that sulfur would outgas as S$_{2}$, much as it does on Io \cite<e.g.,>{Zolotov1999}, and as H$_{2}$S \cite{Renggli2017}. Additionally, elemental sulfur rapidly polymerizes to form aerosol molecules, and has a negligible saturation vapor pressure for the temperatures shown here \cite{Lyons2008,Harman2018}. Given that the S flux is comparable to the flux of CO, removing some or most of the S as a secondary species or aerosol would reduce the mass of the atmosphere, leaving CO and H$_{2}$ as the predominant contributors to the lunar atmosphere \cite{Renggli2017}. We can take this one step further and introduce the fluxes of volcanic gases into a box model for atmospheric chemistry (not shown, but it is based on the reducing chemistry scheme seen in \citeA{Harman2015}), assuming a range of temperatures potentially relevant to the Moon and the appropriate solar spectrum for a $\sim$600 My-old Sun \cite{Claire2012}. Initial tests at 10 mbar suggest that the predominant species in the lunar atmosphere at steady state are dependent upon the temperature, with low temperatures ($\lessapprox$150 K) preserving CO from conversion to CO$_{2}$ via chemical interaction with water vapor photolysis products (namely OH), and modest temperatures ($\gtrapprox$175 K) conversely promoting a CO$_{2}$-dominated atmosphere. 
This is a known phenomenon for cold CO$_{2}$-dominated atmospheres \cite{Zahnle2008,Gao2015}. We have not interactively resolved the chemistry in the 3-D model, but have instead chosen to simulate two end-member scenarios where the atmosphere is either primarily CO or CO$_{2}$. However, it is important to note that while the conversion between CO and CO$_{2}$ is mediated largely by OH, a source of oxygen is also necessary. This oxygen could be produced through the loss of hydrogen to space (derived from water vapor photolysis; see Section \ref{escape}). 

\section{Atmospheric loss processes}\label{escape}

Thermal escape from a modest lunar atmosphere could substantially modify not only the chemical composition, but potentially the total atmospheric mass if escape rates are large enough. NK2017 suggest thermal escape rates on the order of 10 kg s$^{-1}$ based on earlier estimates \cite{Vondrak1974,Vondrak1974a}. However, the total atmospheric mass of a 10-mb lunar atmosphere would be $\sim2\times10^{16}$ kg, rather than the $\sim10^{9}$ kg suggested by \citeA{Vondrak1974}. This large difference in mass changes the altitude of the exobase, and therefore increases the surface across which most of the escape occurs. 

There exist several potential limits to the escape flux, but here we focus on escape rates in which H is diffusion-limited \cite<see>{Catling2017}, assuming a CO$_{2}$-dominated atmosphere with less-abundant lighter gases, such as H$_{2}$ and CO, as discussed in Section \ref{chemistry}. For these calculations, the escape parameter (or Jeans parameter) ($\lambda = \frac{GM_{p}m}{kT_{exo}(R_{p}+h_{exo})}$) is useful to evaluate the stability of the atmosphere and indicates the thermal escape regime. Here, $M_{p}$ is the lunar mass,  $R_{p}$ is the planet radius, $h_{exo}$ is the height of the exobase, $m$ is the mass of the escaping atom or molecule, and $T_{exo}$ is the exospheric temperature. For large values of $\lambda$, the atmosphere is hydrostatic, and only molecules with sufficient energy will escape (Jeans escape) \cite{Catling2017}. When $\lambda \lessapprox 2-3$, the escape becomes hydrodynamic. Based on estimates for the early Earth, Venus, and Mars, we might expect the early Moon's exosphere to be somewhere between 300--1,000 K if it was CO$_{2}$-dominated, and warmer if not \cite{Kulikov2006,Kulikov2007}. This is turn suggests that some species may be escaping hydrodynamically.

In Table \ref{table:Escape} Jeans escape rates are given for each species as a function of the thermospheric/exospheric temperature at the Moon \cite{Vondrak1974a} which is affected by a number of factors\cite{Fahr1983,Zhang1993}. We assume an isothermal atmosphere, so the exospheric altitude varies with the temperature. We consider the exospheric altitude for a CO$_{2}$ atmosphere in all cases, with a ground temperature of 250 K and the temperature in the atmosphere equal to the exospheric temperature. We noted hydrodynamic escape conditions with an $h$; the Jeans computations are not valid in these cases. The values in the table suggest that H and H$_2$ would escape hydrodynamically \cite<e.g.,>{Tian2009,Volkov2011}, and potentially drag along heavier species if their quantity in the upper atmosphere were not limited by the photodissociation of H$_2$O and by diffusion through the lower layers of the atmosphere. It is important to note that these assumptions are pessimistic: the lower atmosphere is more likely to be colder on average, which notably reduces the exospheric altitude and therefore the surface area over which escape occurs, lowering the integrated loss rate. Colder temperatures would also limit the water vapor available to photodissociate.

The rate of supply of light gases to high altitudes (i.e. H, H$_{2}$, O, and H$_{2}$O) will restrict the escape rate as the heavy background gas (either CO or CO$_{2}$) will remain largely hydrostatic \cite <see>[also Table \ref{table:Escape}]{Catling2017}. Water would be restricted to the lower atmosphere largely by the temperature, as mentioned previously. For a CO-dominated atmosphere, the diffusion limit for H$_{2}$ is $\Phi_{d} \approx 2 \times 10^{10}\times f_{H_{tot}}$ cm$^{-2} s^{-1}$, or $<6$ kg s$^{-1}$ (assuming that CO behaves similarly to N$_{2}$ as a background gas), and about 50\% higher for a CO$_{2}$-dominated atmosphere \cite{Hunten1973}. The availability of hydrogen-bearing species in the upper atmosphere is represented by the weighted sum of mixing ratios for those species, $f_{H_{tot}}$ \cite{Catling2017}. The diffusion limit is relatively insensitive to temperature \cite<see Section 3.2.2 of>[]{Zahnle1990}.

An important point from Table~\ref{table:Escape} is that O can have a large escape rate, even for relatively cold exospheric temperatures. However, such large escape rates are not realistic since O comes from CO$_2$ or CO dissociation. The dissociation rate of CO and CO$_2$ as computed by Aeroplanets \cite{Gronoff2012a, Gronoff2012b, Gronoff2014} in a young Sun scenario does not reach that magnitude, leading to a dissociation rate-limited scenario that should be investigated further.  

\begin{table}[ht!]
\caption{Atmospheric Jeans Escape Rates}
\label{table:Escape}
\centering
\begin{tabular}{|l|c|c|c|c|c|c|c|c|c} 
\hline
Exospheric Temperature & \multicolumn{2}{c|}{500 K}  & \multicolumn{2}{c|}{700 K} & \multicolumn{2}{c|}{1500 K} & \multicolumn{2}{c|}{3000 K}\\
Exospheric Altitude & \multicolumn{2}{c|}{1200km} & \multicolumn{2}{c|}{1721km} & \multicolumn{2}{c|}{3800km} & \multicolumn{2}{c|}{8000km}\\
\hline
Species  & $\lambda$ & $\Phi_{i}$ & $\lambda$ & $\Phi_{i}$ & $\lambda$ & $\Phi_{i}$ & $\lambda$ & $\Phi_{i}$ \\
\hline
CO$_{2}$ & 17  & 0.04   & 10  & 20                 & 3.11 &9.6$\times$10$^{3}$  & 0.9 & $h$\\
CO       & 11  & 7.40   & 7   & 340                & 2    &11$\times$10$^{3}$   & 0.5 & $h$\\
O        & 6.5 & 240    & 4   &1.7$\times$10$^{3}$ & 1.1  & $h$                 & 0.32 & $h$\\
H        & 0.4 & $h$    & 0.2 & $h$                & 0.07 & $h$                 & 0.02 & $h$\\
\hline
\multicolumn{9}{l}{$\lambda$ : Jeans parameter value.}\\
\multicolumn{9}{l}{$\Phi_{i}$ : Jeans Escape rates in units of kg s$^{-1}$.}\\
\multicolumn{9}{l}{$h$ : Full hydrodynamic escape. Jeans computations not valid.}
\end{tabular}
\end{table}

Suprathermal (or non-thermal) escape mechanisms include charge exchange, ion pickup, sputtering and solar wind pickup among other processes \cite{Catling2017} for non-magnetized bodies. A magnetized Moon \cite{Tikooe1700207} would have suffered polar wind escape \cite{Garcia2017}. These loss mechanisms would also contribute to the evolution of the lunar atmosphere by driving off heavier molecules like O, independent (and potentially in excess) of thermal escape \cite<e.g.,>[]{Hunten1982,Airapetian2017} and could ultimately drive the loss of the residual lunar atmosphere. However, models including both avenues have suggested thermal escape would dominate for Earth-sized planets \cite{Kislyakova2013, Erkaev2013}, as well as for the Moon \cite{Vondrak1974}. The Moon would also spend some part of its orbit in the Earth's shadow (see Section \ref{orbital}), which would further reduce the impact of suprathermal escape processes. It is important to note, however, that the Earth's magnetic field would have been compressed by more frequent coronal mass ejections \cite{Airapetian2015,Airapetian2016}. As an aside, the early Moon may have experienced a large number of impacts, which could have enhanced atmospheric escape \cite{Zahnle2017}; this does not fall neatly into either the thermal or suprathermal loss categories. 
Since the O loss is limited by the dissociation of CO and CO$_2$, the suprathermal escape only increase the losses marginally. It is therefore possible to estimate the loss rate of a 700 K thermosphere composed of 99.5\% CO$_2$ and 0.5\% O (analogous to the Martian atmosphere) at 30~kg/s. In that case, the outgassing rate from the maria as conjectured by NK2017 is able to compensate for the escape, and allows the accumulation of an atmosphere, paralleling the impacts of the Apollo missions \cite{Vondrak1992}. The main question lies with the ability of the Moon to create an initial collisional atmosphere from an exosphere in these conditions: Io has a much higher creation rate but is not able to accumulate an atmosphere \cite{Koga2018}. Ideally escape rates should be calculated as the atmosphere builds up, but as mentioned above we assume the atmosphere already exists for our calculations.

\section{Model setup}\label{model}

\subsection{Orbital parameters}\label{orbital}

The latest research \cite<e.g.>[]{Canup2004,Halliday2008,Herwartz2014,Lock2016,Lock2017} tends to support the canonical hypothesis of the Moon forming $\sim$4.5 Gya as a result of a Mars-sized impactor colliding with a  proto-Earth \cite{Hartmann1975}, though the hypothesis of multiple impacts has gained popularity in recent years \cite{Rufu2017}.

The ejected material formed a disk that accreted and formed the Moon. The shape of the tidal bulge that formed at the time when Moon had solidified allows one to obtain an estimate for its distance from the Earth at that time \cite<$\sim$14.5 Earth radii, see>[]{Crotts2014}. The time evolution of its orbit to the present value of $\sim$60 Earth radii is less constrained and is still subject to debate. Most research \cite{Williams2000,Webb1982} suggests that the Moon's orbit was evolving rapidly in its early stages with the Moon--Earth distance $\sim$0.75 of its present day value (60 Earth radii) $\sim$0.5 Gy after formation, though, according to \citeA{Siegler2015}, the Moon could have been much closer to Earth (41-34 Earth radii) at the time in question. 

Another important aspect of the Moon's orbit that is poorly constrained is its obliquity with respect to the normal to the ecliptic. The current value ($\sim$1.5$^\circ$) is rather small and doesn't produce much seasonal variation.
But as the Moon's orbit evolved with time it could have had much higher values. According to several studies \cite{Cuk2016,Ward1975,Siegler2011}, while its distance from the Earth was between 30 and 40 Earth radii the Moon underwent a Cassini state transition where its obliquity could have been as high as 50$^\circ$.  

In our current study we assume that the Moon was at a distance of 45 Earth radii from Earth, which implies a sidereal rotation period of 17.8 modern Earth days. At this distance the obliquity should be close to the modern one of $\sim$1.5$^\circ$. Hence, for simplicity in our simulations we set it to zero. But we plan to address the effect of non-zero obliquity in future studies.

\subsection{The General Circulation Model}
For our experiments we use ROCKE-3D, a planetary general circulation model (GCM) developed at the NASA Goddard Institute for Space Studies \cite[,also see Appendix A]{Way2017,Schmidt2014}. We use $4^\circ\times5^\circ$ latitude$\times$longitude resolution ($\sim$150 km spatial resolution) and 40 atmospheric layers with the upper boundary at 10$^{-4}$ of the surface pressure ($\sim$180-330~km atmospheric height in our experiments).

Since we are interested in the period after the Late Heavy Bombardment, corresponding to the period of the most intensive lunar volcanic activity, we assume that the Moon's surface has not undergone significant resurfacing since then. Hence we use modern observational data to describe its surface. For the Moon's topography we use the Lunar Orbiter Laser Altimeter data set \cite{Barker2016}\footnote{https://pds.nasa.gov/ds-view/pds/viewProfile.jsp?dsid=LRO-L-LOLA-4-GDR-V1.0}. For the surface albedo we use a high-resolution narrow-band normal albedo data set \cite{Lucey2014}, which we  scaled to produce a correct broadband Bond albedo \cite{Buratti1996} when averaged over the entire surface of the planet. To represent the  PSRs we used the \citeA{Mazarico2011} dataset.
We treat PSRs as a special type of surface which is not exposed to
direct solar radiation, but in all other respects is similar to the
rest of the planet's surface. Since our resolution doesn't allow us to
resolve each PSR individually, PSRs are represented statistically, as
fractions of grid cells covered by PSRs. Figure A1 (Appendix A) shows the surface topography and PSR fractions on the ROCKE-3D grid. For our resolution the highest PSR fraction is 28.2\%, which still allows us to treat PSRs as subregions of the grid cells, redistributing the direct solar radiation inside the cell accordingly.

\subsection{Experiments}\label{experiments}

We conducted experiments for 10 mb, 2.5 mb and 1 mb atmosphere as shown in Table \ref{table:Experiments}. They represent the limiting cases for the lunar atmosphere 3.5 Gya as was discussed in previous sections. In all cases the simulations were run until the model reached an equilibrium state, which for a cool dry terrestrial world happens relatively quickly (on the order of 10--20 simulation years). 
We considered cases of ``dry" and ``wet" atmospheres. In the former case the atmosphere was initialized with 10$^{-7}$ kg H$_2$O per kg of atmosphere and stayed close to this humidity level throughout the simulations. In latter case it was initialized to 5$\times$10$^{-3}$ kg kg$^{-1}$ (except for the upper 10\% of the atmosphere, where it was initialized to zero) and evolved to equilibrium together with the rest of the system. One should notice that here we are talking about the meteorological equilibrium. Some amount of water is being lost due to continuous freezing out at the poles and in permanently shadowed regions. But these are very slow processes (once the initial water has settled as frost in polar regions and deep in the soil elsewhere, there is very little supply to the atmosphere), which happen on geological rather than meteorological scales and compete with water delivery by volcanic outgassing and impactors. It is beyond the scope of this work to consider such processes.

\begin{table}[ht!]
\caption{Experiments}
\label{table:Experiments}
\begin{tabular}{|l|l|l|l|l|l|l|} 
\hline
Number& distance     & rotation& obliquity  & atmospheric& surface & initial H$_2$O \\
      & from Earth   & period  & to Sun     & composition& pressure& kg kg$^{-1}$ \\
      & (Earth radii)& (days)  & $^\circ$   &            & (mb)    & atm.\\
\hline
1a,b,c & 45.   & 17.8  & 0   & 100\% CO     & 10, 2.5, 1 & 10$^{-7}$ \\
2a,b,c & 45.   & 17.8  & 0   & 100\% CO$_2$ & 10, 2.5, 1 & 10$^{-7}$ \\
3a,b,c & 45.   & 17.8  & 0   & 100\% CO     & 10, 2.5, 1 & 5$\times$10$^{-3}$ \\
4a,b,c & 45.   & 17.8  & 0   & 100\% CO$_2$ & 10, 2.5, 1 & 5$\times$10$^{-3}$ \\
\hline
\end{tabular}
\end{table}

\section{Results and Discussion}\label{discussion}

\begin{figure}
\includegraphics[scale=1.]{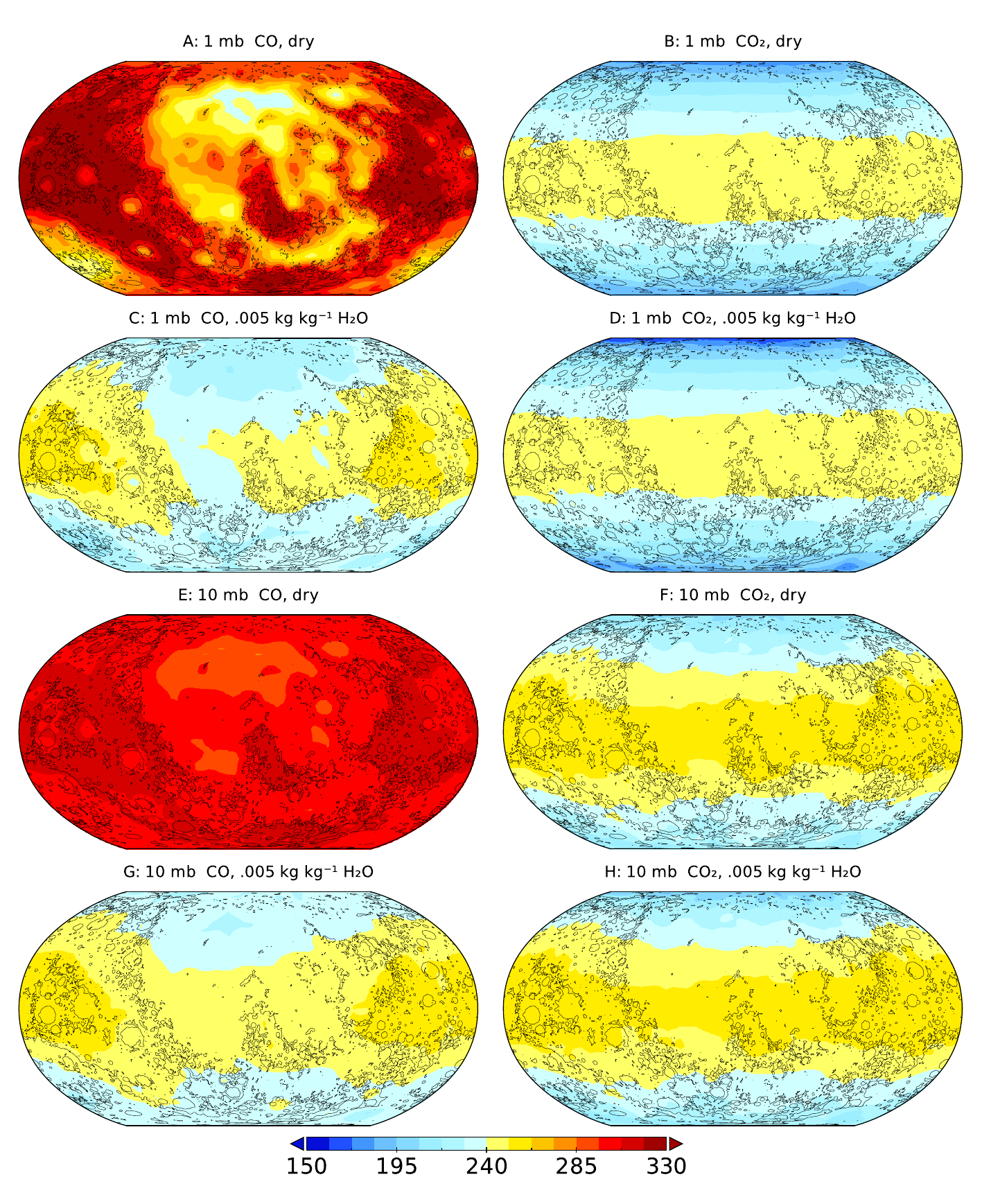}

\caption{\small Lower atmospheric layer (2\% mass) temperature (K) for a pure CO (left) and a pure CO$_2$ (right) 1~mb (A,B,C,D) and 10~mb (E,F,G,H) atmosphere. The upper row represents the experiments with a dry atmosphere (experiments 1c and 2c in Table~\ref{table:Experiments}). The second row corresponds to experiments that were initialized with 0.005 kg H$_2$O per kg of atmosphere (experiments 3c and 4c in Table~\ref{table:Experiments}). The experiments in the third and forth rows are similar to the ones in the top rows, except that they correspond to 10~mb atmosphere (experiments 1a, 2a, 3a and 4a in Table~\ref{table:Experiments}).} \label{FigTatm1}
\end{figure}

\begin{figure}
\includegraphics[scale=0.6]{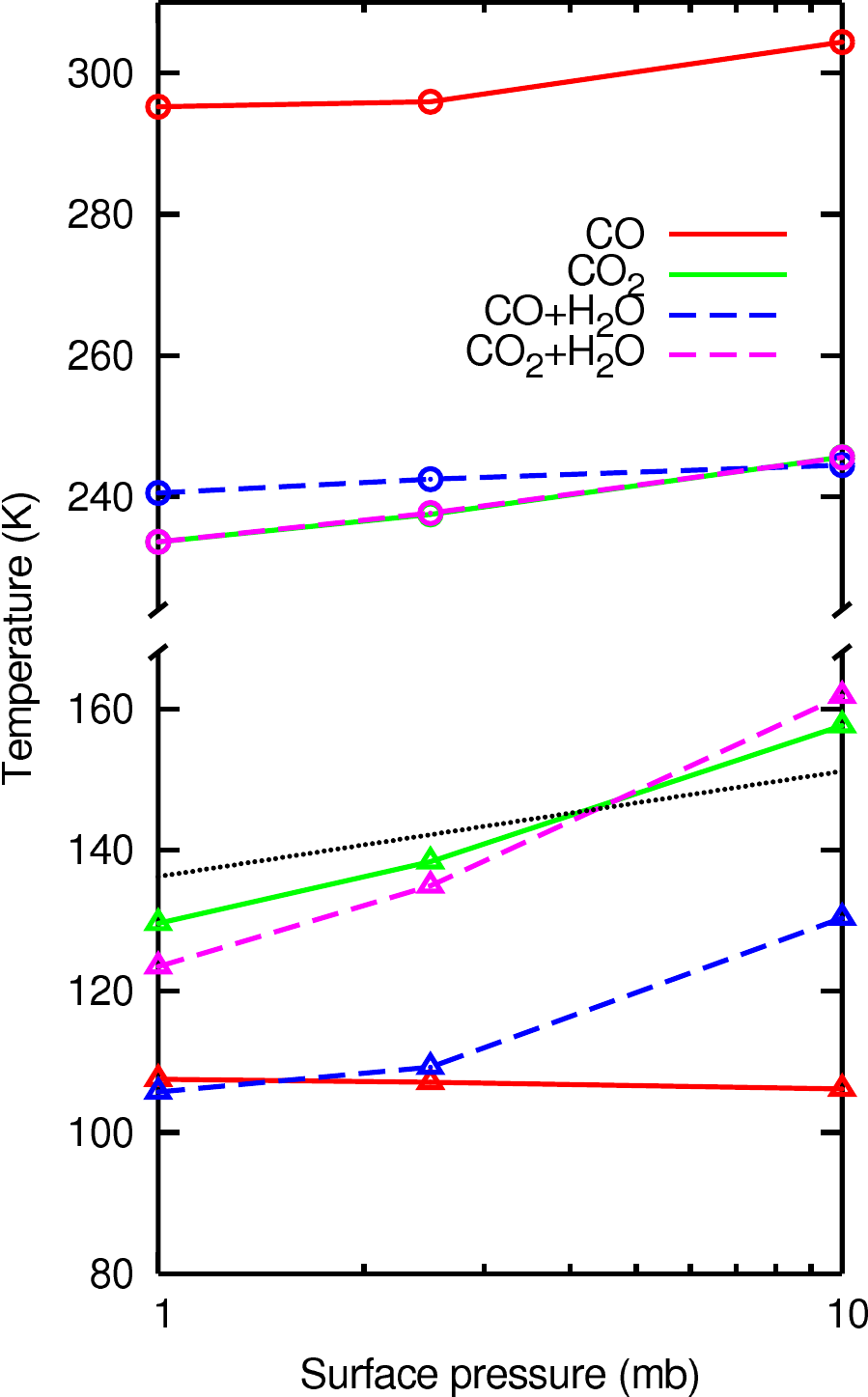}
\centering
\caption{\small Global mean temperature of the lower layer (2\% mass) of the atmosphere (circles) and ground temperature at the poles (triangles) as a function of atmospheric thickness. The marks show the results of experiments for 1, 2.5 and 10 mb atmosphere. Black dotted line represents the CO$_2$ condensation curve.} \label{FigTplot}
\end{figure}

The distinctive property of a climate on a slowly rotating planet with
a thin atmosphere is a very strong diurnal cycle. In all our
experiments the typical ground temperature at night was $\sim$200 K,
while during the day at the sub-stellar point it was as high as
$\sim$340 K. Figures A3, A4 and A5 (Appendix A) show instantaneous maps of ground temperature, lower atmosphere temperature and surface pressure for 1 mb experiments. While in the CO$_2$ case the atmospheric temperature basically follows the ground temperature, the CO case is more complicated, with more uniform temperatures over the surface and the effect of topography more pronounced. We discuss this in more detail below, when we look at Figure~{\ref{FigTatm1}}.
Despite cold temperatures during the lunar night, in all our experiments a diurnal cycle ensures that any water cold-trapped on the night side only remains there temporarily.  Permanent cold trapping is only possible at the poles, where the temperatures remain consistently cold.

Figure \ref{FigTatm1} (A,B,C,D) shows maps of annually-averaged air temperature of the lowest atmospheric layer (constituting 2\% of the entire atmospheric mass) for the 1 mb GCM simulations presented in Table \ref{table:Experiments}. Of particular interest is to compare plots A and B, that correspond to CO and CO$_2$ atmospheres on a dry planet. At first glance the results seem to be counter-intuitive as the greenhouse gas dominated atmosphere (CO$_2$) is for the most part cooler than the radiatively neutral (CO) atmosphere. This happens not only at the surface, but throughout the entire depth of the atmosphere. The average bulk atmospheric temperature is 211.3 K for the CO$_2$ atmosphere and 281 K for the CO atmosphere. The explanation comes from the fact that in the case of a slowly rotating planet with such a thin atmosphere a significant amount of atmospheric heating is received from the ground through turbulent heat flux during the daytime when the ground is hot. During the night time the ground cools quickly through radiative cooling which leads to stable stratification of the planetary boundary layer and effective suppression of turbulent exchange fluxes between the ground and atmosphere. So, the only possibility for the atmosphere to cool during the night time is through radiative cooling. But a radiatively neutral CO atmosphere cannot radiatively cool, basically preserving the energy obtained from the ground during the daytime and staying warmer than the CO$_{2}$-dominated ``greenhouse" atmosphere. 

When one adds small amounts of water to the system, as we have done in experiments C and D, the CO-dominated atmosphere is no longer radiatively neutral (water vapor being a greenhouse gas), and the results become more conventional. While adding water has very little effect on the CO$_2$-dominated atmosphere (the bulk temperature increased slightly to 216 K), in the CO case the surface air temperature and the entire atmosphere become much colder, with average bulk atmospheric temperature being 225.9~K.

The average equilibrium amount of water in lower atmosphere was 6$\times$10$^{-4 }$ kg/kg and 1$\times$10$^{-4 }$ kg/kg in CO and CO$_2$ 1~mb ``wet'' experiments respectively. Though water content was higher in the CO atmosphere, we found that its sequestration rate to PSRs was lower as a result of extremely stable stratification due to much lower temperatures inside the PSRs (Figure A2). Such low PSR temperatures in the CO case are due to the lack of greenhouse heating present in the CO$_{2}$ atmosphere.

Plots E,F,G,H in Figure {\ref{FigTatm1}} show temperature maps for the 10~mb atmospheres. While most of these experiments exhibit similar behavior to the 1~mb case, the dry CO atmosphere shows a more even distribution of temperature, most likely due to reduced topographic effect (the relative surface pressure difference between low and high altitude is lower for a thicker atmosphere.

Figure~\ref{FigTplot} presents the global average temperature of the first atmospheric layer for all 12 experiments listed in Table~\ref{table:Experiments} (marked by circles). As in the 1 mb case, the CO$_2$ dominated atmosphere ``dry" and ``wet" experiments basically coincide, while for the CO atmosphere they are separated by $\sim$50 K. In all experiments the temperature increases as the atmosphere gets thicker, due to the increased greenhouse effect and more efficient turbulent transfer of the heat from the surface. One should notice that in all cases the average atmospheric temperature is much warmer than the threshold discussed in Section~\ref{chemistry}, so provided that one has a sufficient water supply it is likely that the atmosphere would be CO$_{2}$-dominated. 

Figure \ref{FigTplot} also presents the ground (regolith) temperature at the north pole (those experiments are marked with triangles). This is the coldest temperature at the surface of the Moon outside the PSRs. As one would expect, this temperature also increases with increasing atmospheric thickness (due to the greenhouse effect), except for the case of the dry CO atmosphere, which is radiatively neutral (the slight negative trend is due to increased Rayleigh scattering in thicker atmospheres). One important fact to notice is that the lines representing the experiments with a CO$_2$-dominated atmosphere cross the CO$_2$ condensation threshold (black dotted line). Hence, for surface pressures of 2.5 mb and lower such an atmosphere is prone to collapse, condensing at the poles. Condensation will be even more intense inside PSRs. Our estimates give that the 1 mb CO$_2$-dominated atmosphere will lose about 0.01 of its mass per year due to condensation around the poles and in PSRs. The CO-dominated atmosphere doesn't have this problem, since the CO condensation temperature is much lower ($\sim$50 K at 1 mb pressure). Extremely cold temperatures at the poles are partially a consequence of the zero obliquity used. If the Moon had a substantial obliquity at the epoch modeled herein it could warm the poles and mitigate the problem of a potential collapse of the CO$_2$-dominated atmosphere.

The past outgassing rates from lunar maria are a subject of active research and are still poorly constrained. In this work we used typical atmospheric pressures which would result from the relatively high outgassing rates conjectured by NK2017. In the future we plan to consider thinner atmospheres, which would result from lower outgassing rates proposed in other research. More work can be done to estimate atmospheric escape rates by simulating a realistic upper atmosphere. This could lead to the estimation of an additional deuteration in the water potentially deposited at the poles, which would be a way to confirm the existence of this past atmosphere.

%
%
%
%
%
%
%
%

\acknowledgments
This work was supported by NASA's Nexus for Exoplanet System Science (NExSS). Resources supporting this work were provided by the NASA High-End Computing (HEC) Program through the NASA Center for Climate Simulation (NCCS) at Goddard Space Flight Center.
The work of GG is supported by the NASA Astrobiology Institute grant NNX15AE05G. 
This work benefited greatly from discussions with Jim Head and Christopher Hamilton as well as our colleagues David Rind and Max Kelley.
The data used to generate Figures can be downloaded from the NCCS data portal: \\
https://portal.nccs.nasa.gov/GISS\_modelE/ROCKE-3D/publication-supplements/

\bibliography{references}
%

\clearpage

\appendix 
\section{Supporting Information}

\noindent\textbf{Radiative transfer model configuration}

For radiative transfer ROCKE-3D employes SOCRATES radiation model  \cite{Edwards1996a,Edwards1996b,Way2017}.
As described in section 2, it is likely that the early lunar atmosphere evolved as either a CO- or CO$_{2}$-dominated atmosphere.  Here, we leverage existing\footnote{https://simplex.giss.nasa.gov/gcm/ROCKE-3D/Spectralfiles.html} SOCRATES gas absorption coefficient tables (known as spectral files) for use in our 3D simulations.  A CO$_{2}$-dominated early lunar atmosphere closely resembles the Martian atmosphere, and thus existing spectral files originally constructed for the planet Mars are applied.  However, for the CO-dominated atmospheres modeled herein we have used existing spectral files originally constructed for N$_{2}$-dominated atmospheres.
This is possible because CO and N$_{2}$ act similarly as broadening and scattering gases, but have limited impact on absorption in the atmosphere. 

It is also convenient that CO and N$_{2}$ have nearly identical molecular weights and specific heats.  
Using a 1-D offline version of SOCRATES we performed sensitivity tests to quantify differences in the radiative transfer between CO and N$_{2}$ dominated atmospheres.  For this 1-D testing we use a nominal early Moon profile consisting of 10\% CO$_{2}$, $\sim$1\% H$_{2}$O, and either CO or N$_{2}$ as the background gas constituting the rest of the atmosphere.  We assume a surface pressure of 9.28 mb, a surface temperature of 250 K, falling to 175 K and becoming isothermal at $\sim$0.006 mb and lower pressures.   

In the longwave, N$_{2}$ only effects the radiative balance via N$_{2}$-N$_{2}$ collision induced absorption (CIA) beyond $\sim$20 $\mu$m \cite{WordsworthPierrehumbert2013}, however its effect is negligible on the early Moon given the paucity of the atmosphere.  CO has weak absorption features between 4.5 -- 4.8 $\mu$m and also beyond $\sim$50 $\mu$m \cite{Wordworth2015}.  However, these absorption features lie on the shoulder of the Planck distribution of thermal emission from our nominal early Moon atmosphere, and generally are overlapped by H$_{2}$O absorption.  In 1-D SOCRATES calculations, we find that there is only $\sim$0.3 Wm$^{-2}$ difference in the outgoing longwave radiation at the top of the atmosphere between N$_{2}$ dominated and CO dominated atmospheres for the early Moon.  The presence of any CO$_{2}$ and H$_{2}$O will dominate the thermal radiation budget.

In the shortwave, N$_{2}$ has no meaningful absorption features, but CO has weak absorption features in the near-infrared that extend to $\sim$1.2 $\mu$m.  CO also has Rayleigh scattering coefficients that are $\sim$1.37 times greater than those of N$_{2}$.  Combining the two effects, in 1-D SOCRATES calculations we find that an N$_{2}$ dominated atmosphere permits up to $\sim$2 Wm$^{-2}$ more solar radiation to reach the lunar surface, compared with a CO dominated atmosphere.  The paucity of the early lunar atmosphere means that it is largely transparent to both longwave and shortwave radiation, thus attenuation of the incoming solar and outgoing longwave radiation is generally only 10 to 20 Wm$^{-2}$ (i.e. on the order of 1 to 2\% depending on the solar zenith angle).  Thus, the surface albedo dominates the shortwave radiation budget.  Shortwave heating rates in the atmosphere are dominated by near-infrared absorption by water vapor, when water is present in any amount.


\begin{figure*}[ht!]
\includegraphics[trim={0 2cm 0 0},clip,scale=0.28]{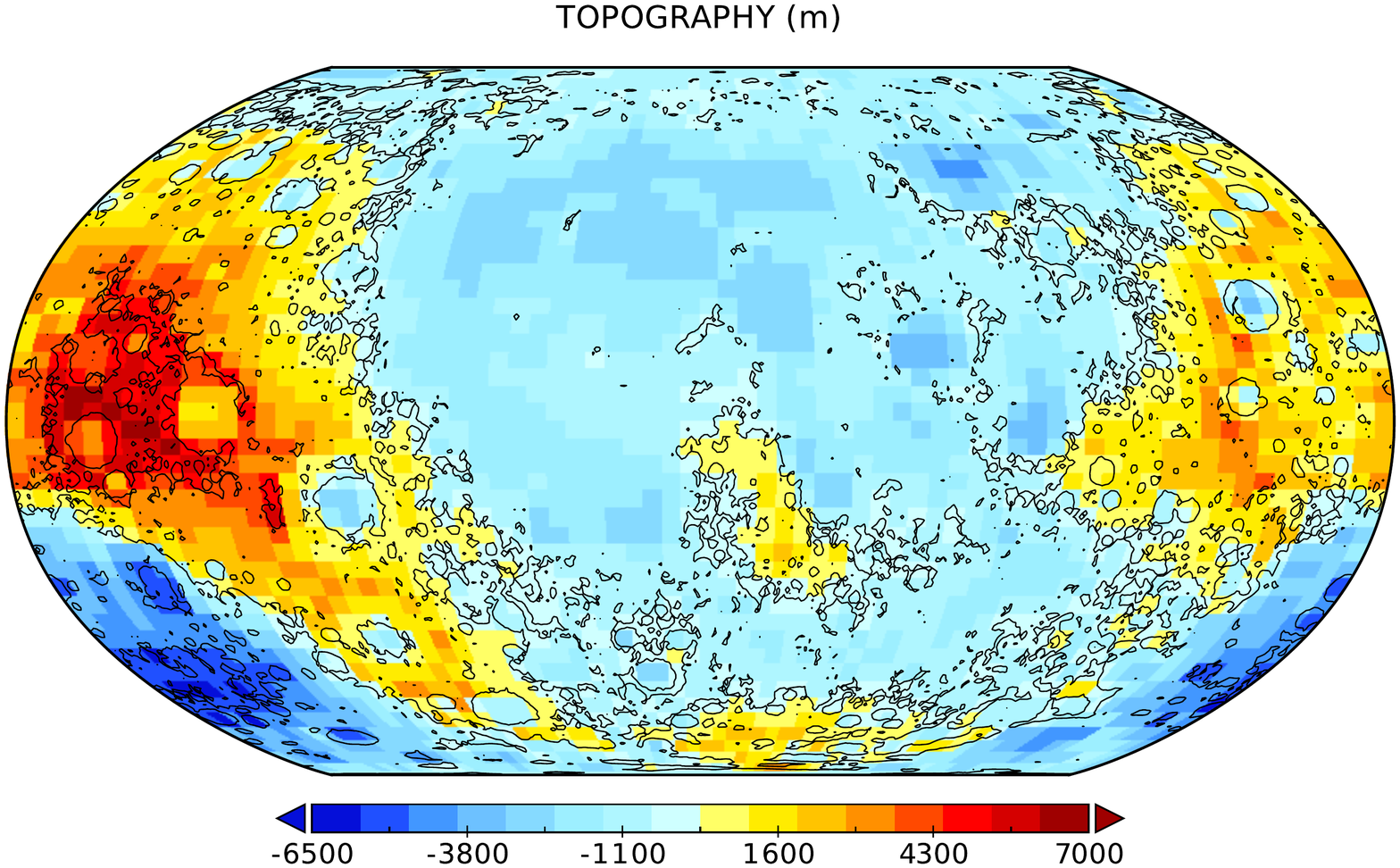}
\includegraphics[trim={0 2cm 0 0},clip,scale=0.28]{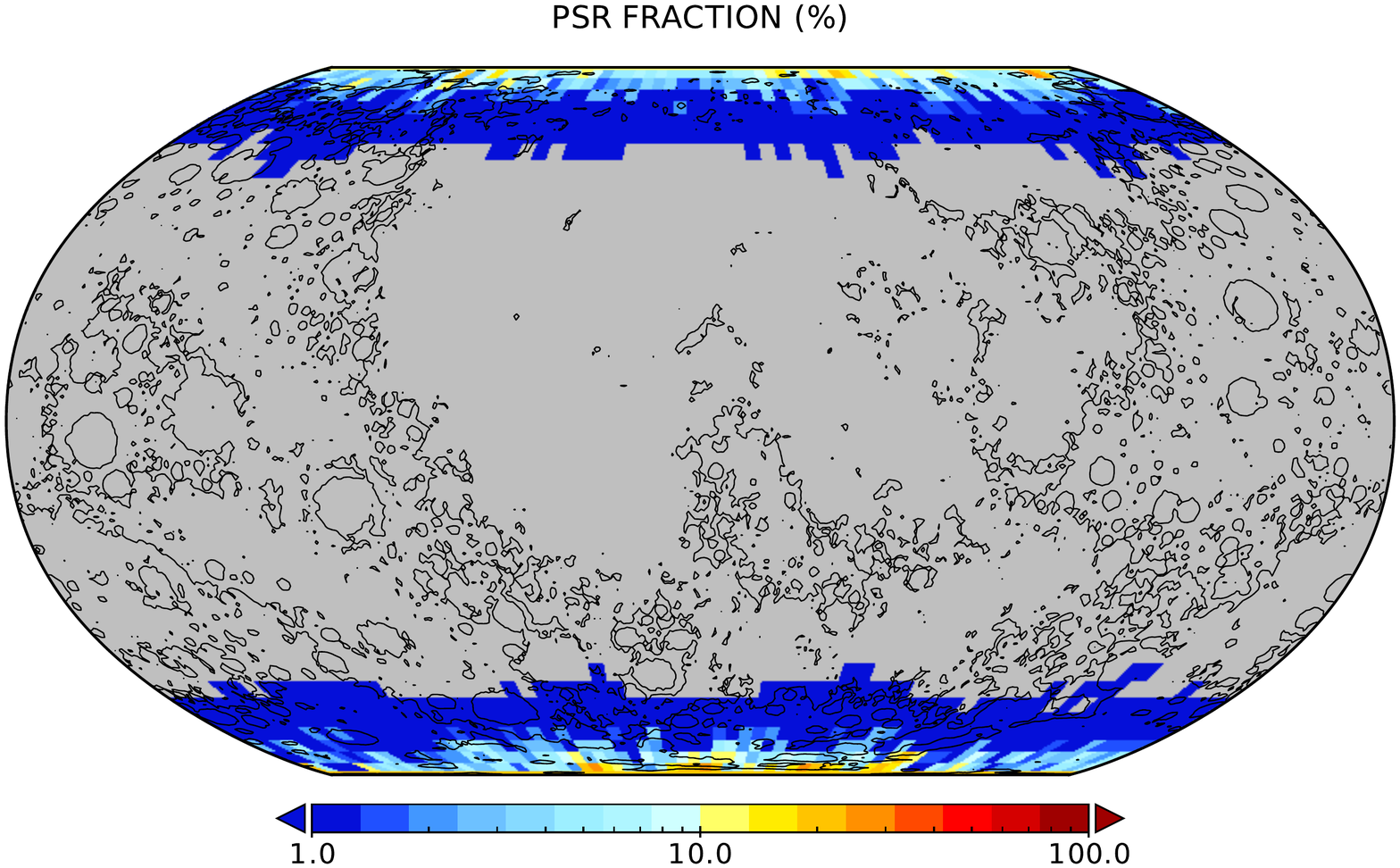}

\caption{\small Moon topography and PSR fractions used as boundary conditions by ROCKE-3D. Fractions of PSRs are shown on logarithmic scale. Maximum PSR fraction in current simulations was 28.2\%. Gray area on the map represents cells with PSR fraction less than 1\% or no PSRs.  }.
\end{figure*}


\begin{figure*}[ht!]
\includegraphics[trim={0 2cm 0 0},clip,scale=0.28]{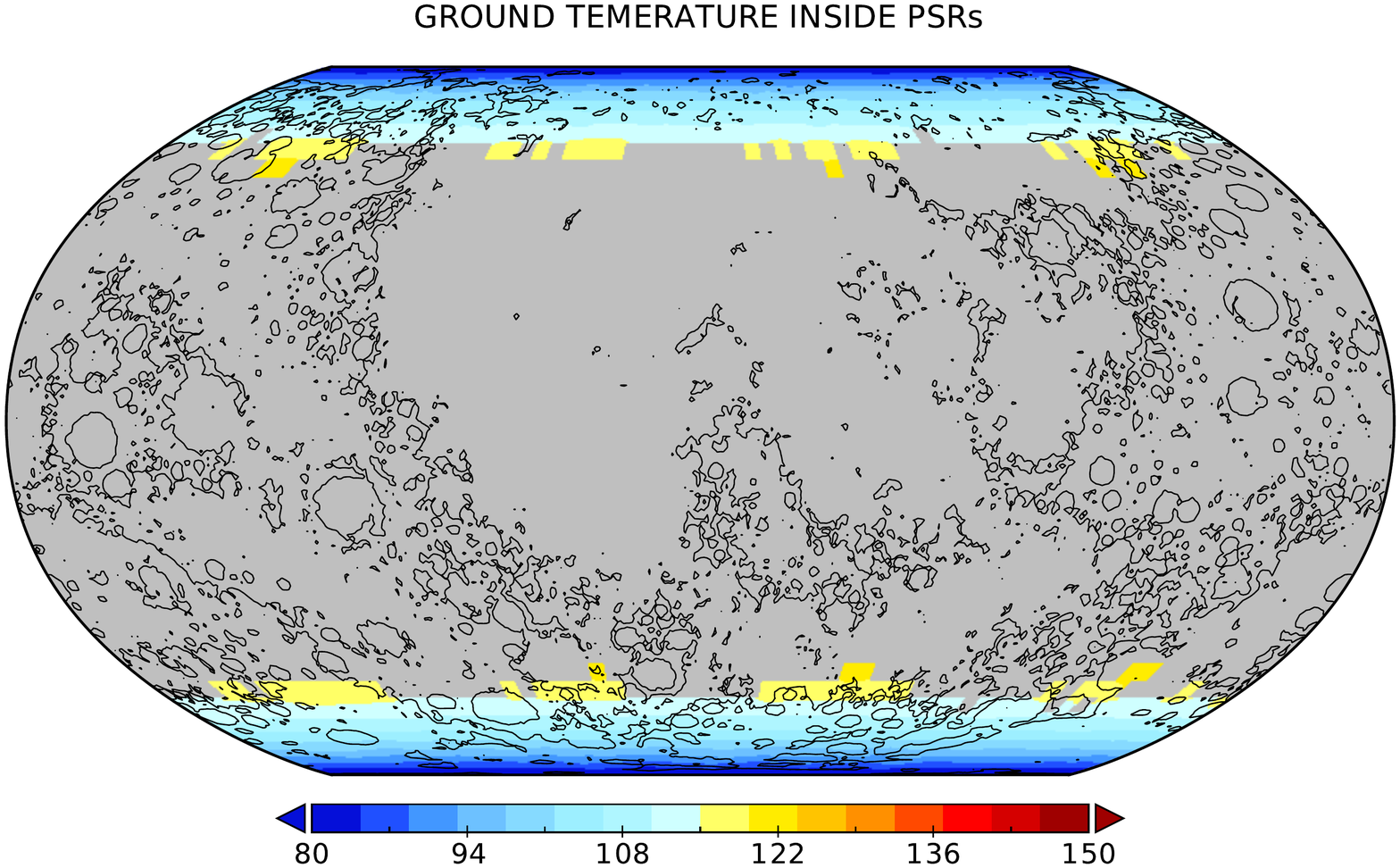}
\includegraphics[trim={0 2cm 0 0},clip,scale=0.28]{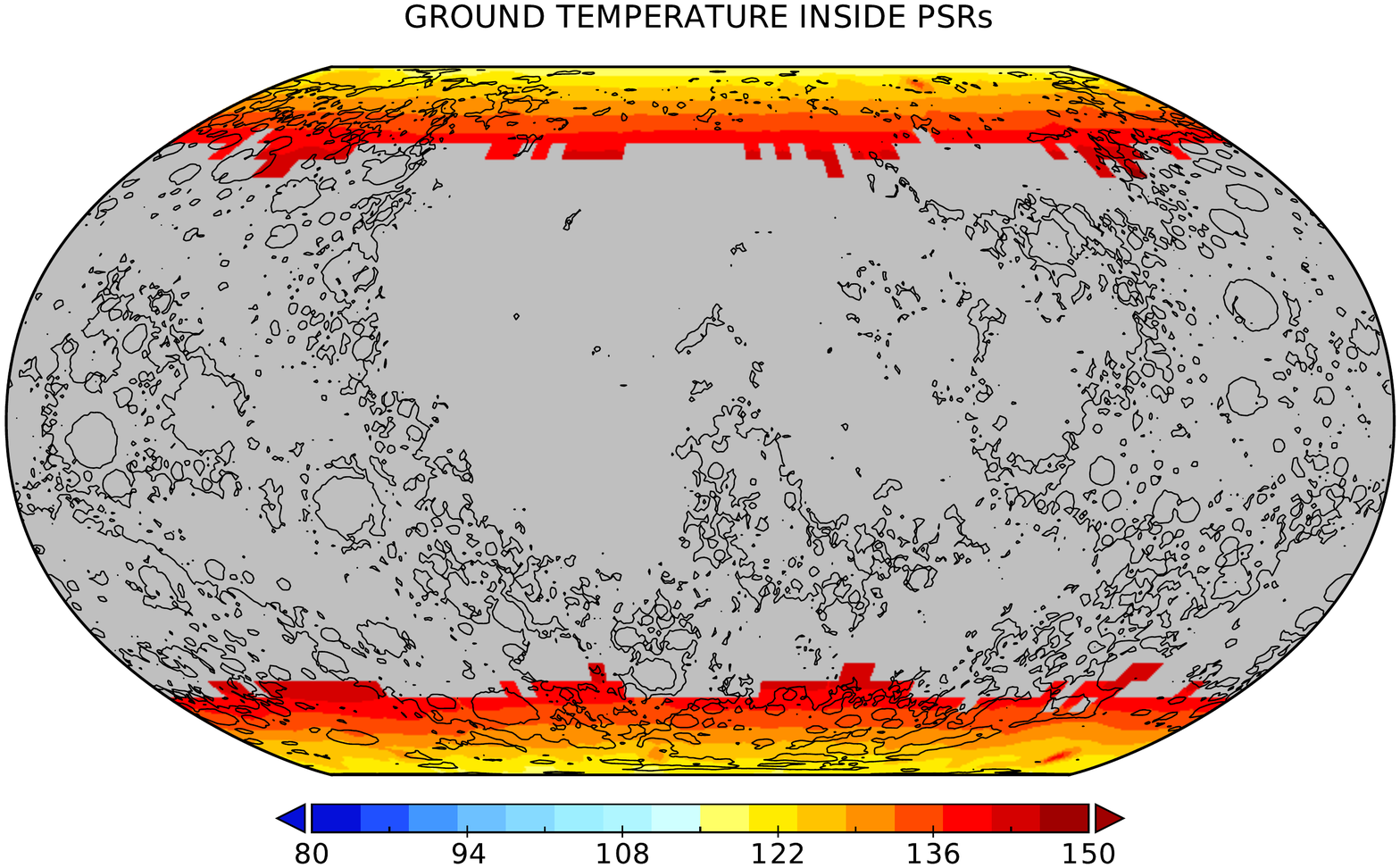}

\caption{\small Ground temperature (K) inside the PSRs for 1 mb ``wet'' CO (left) and ``wet'' CO$_2$ (right) atmosphere. (Experiments 3c and 4c from Table 2.)}.
\end{figure*}


\begin{figure*}[ht!]
.\\
\includegraphics[trim={0 2cm 0 0},clip,scale=0.28]{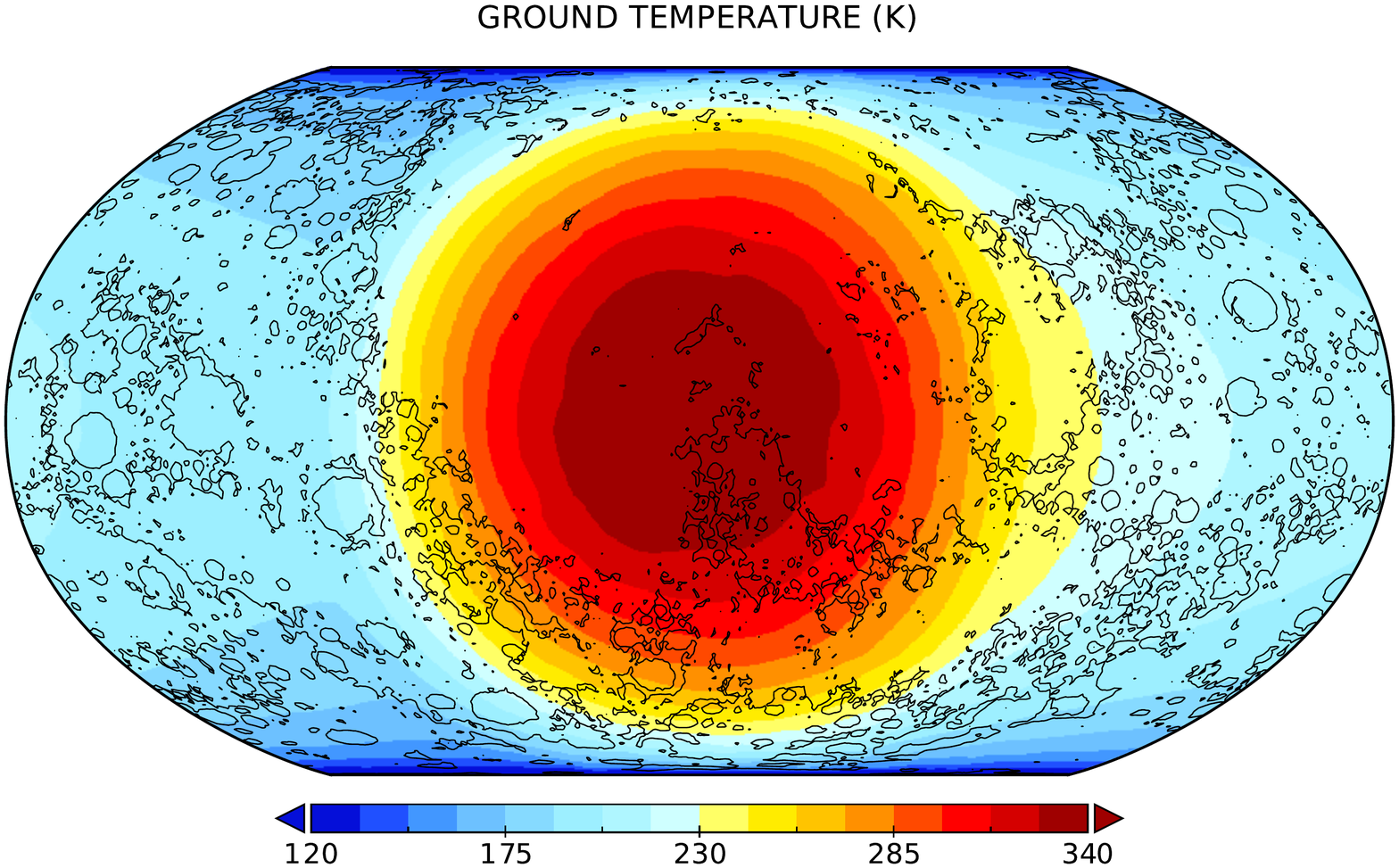}
\includegraphics[trim={0 2cm 0 0},clip,scale=0.28]{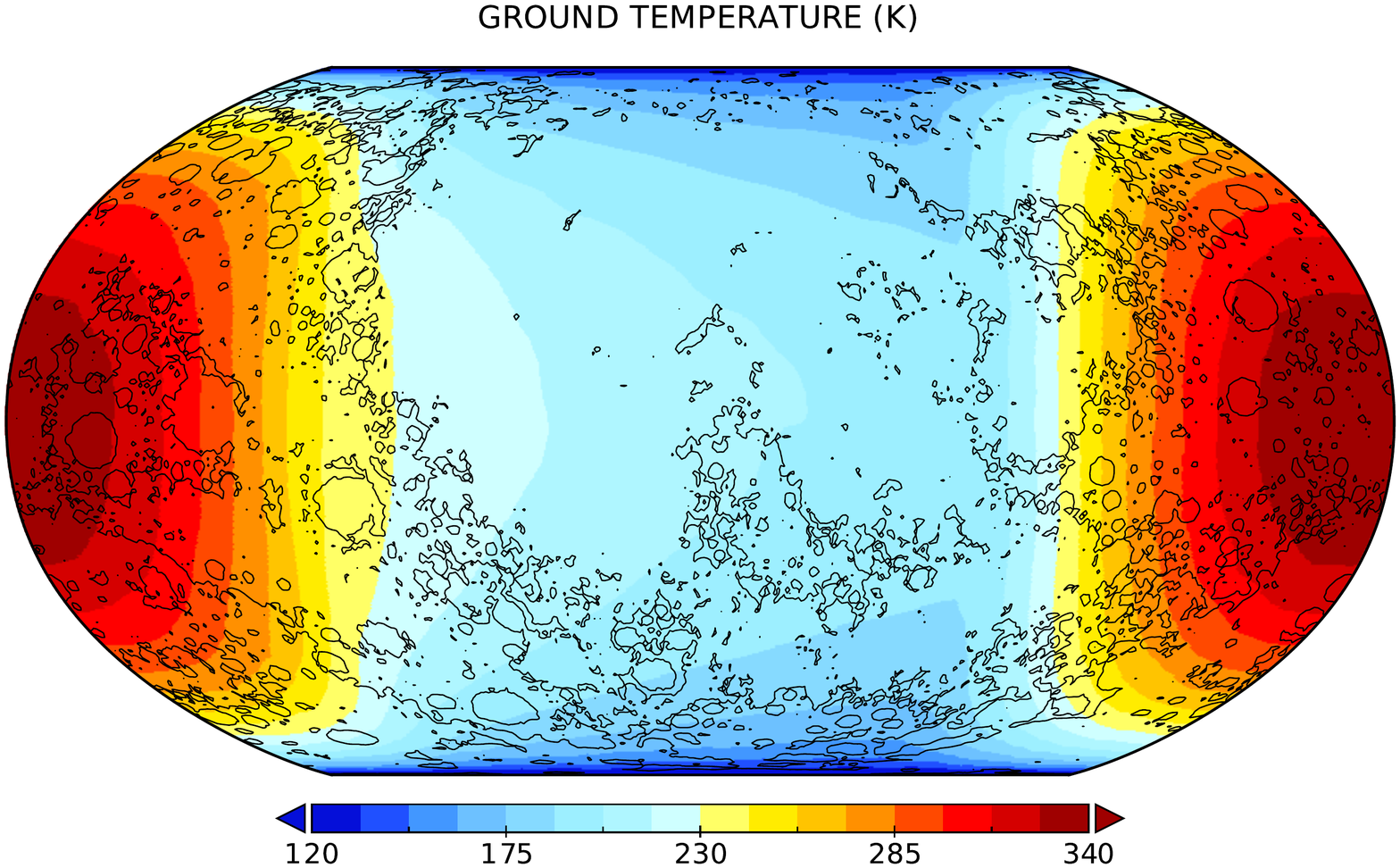}\\
\includegraphics[trim={0 2cm 0 0},clip,scale=0.28]{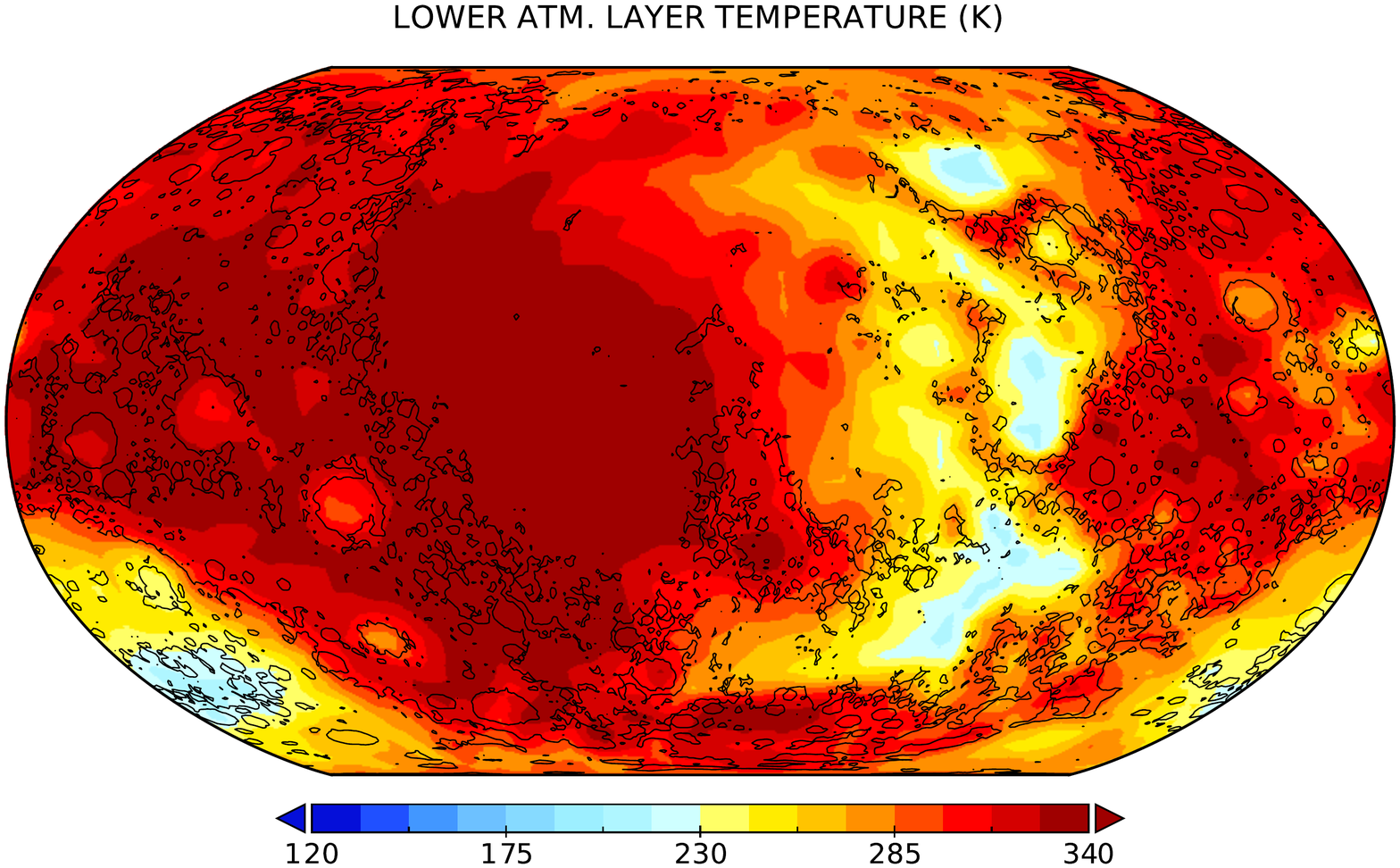}
\includegraphics[trim={0 2cm 0 0},clip,scale=0.28]{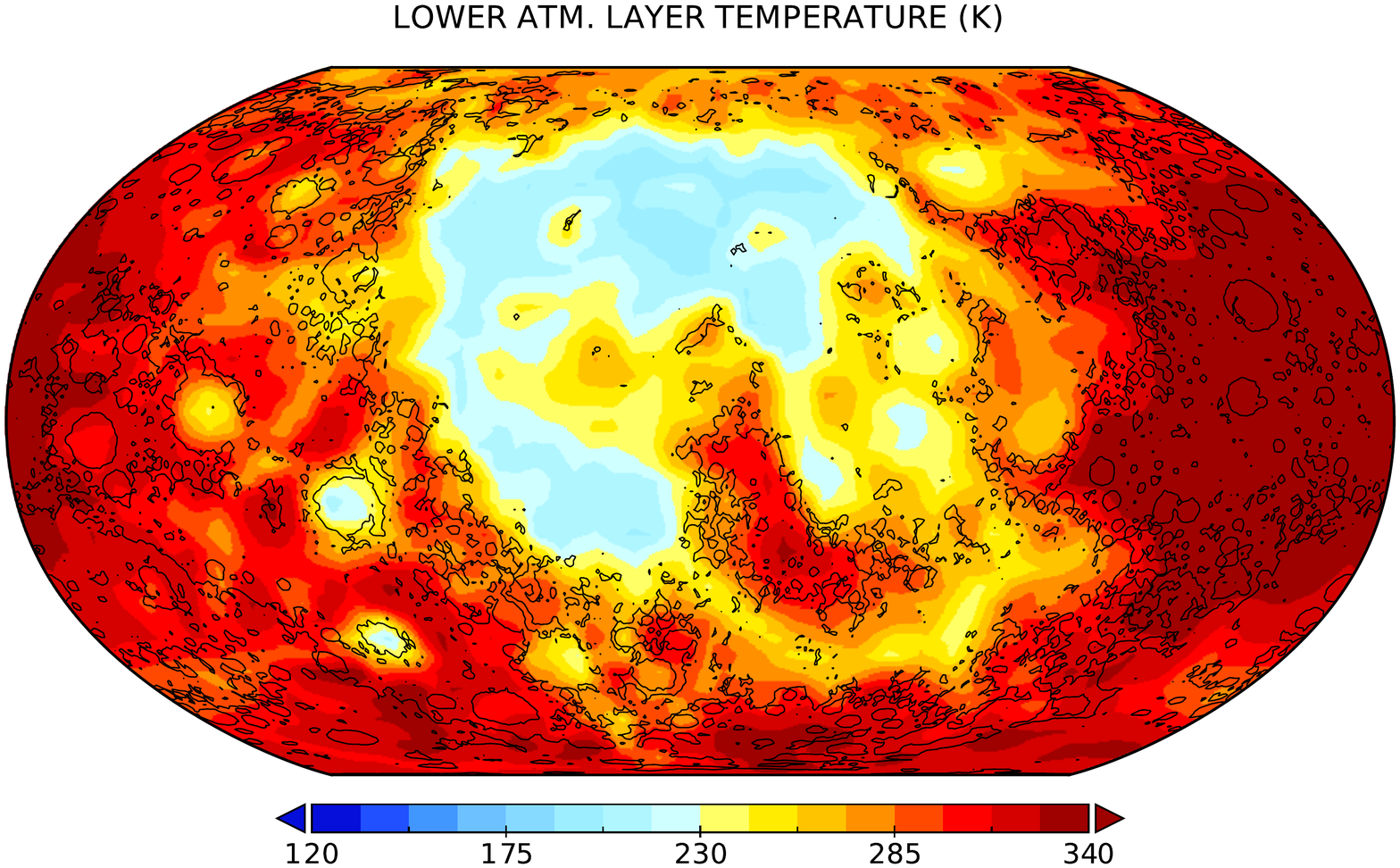}\\
\includegraphics[trim={0 2cm 0 0},clip,scale=0.28]{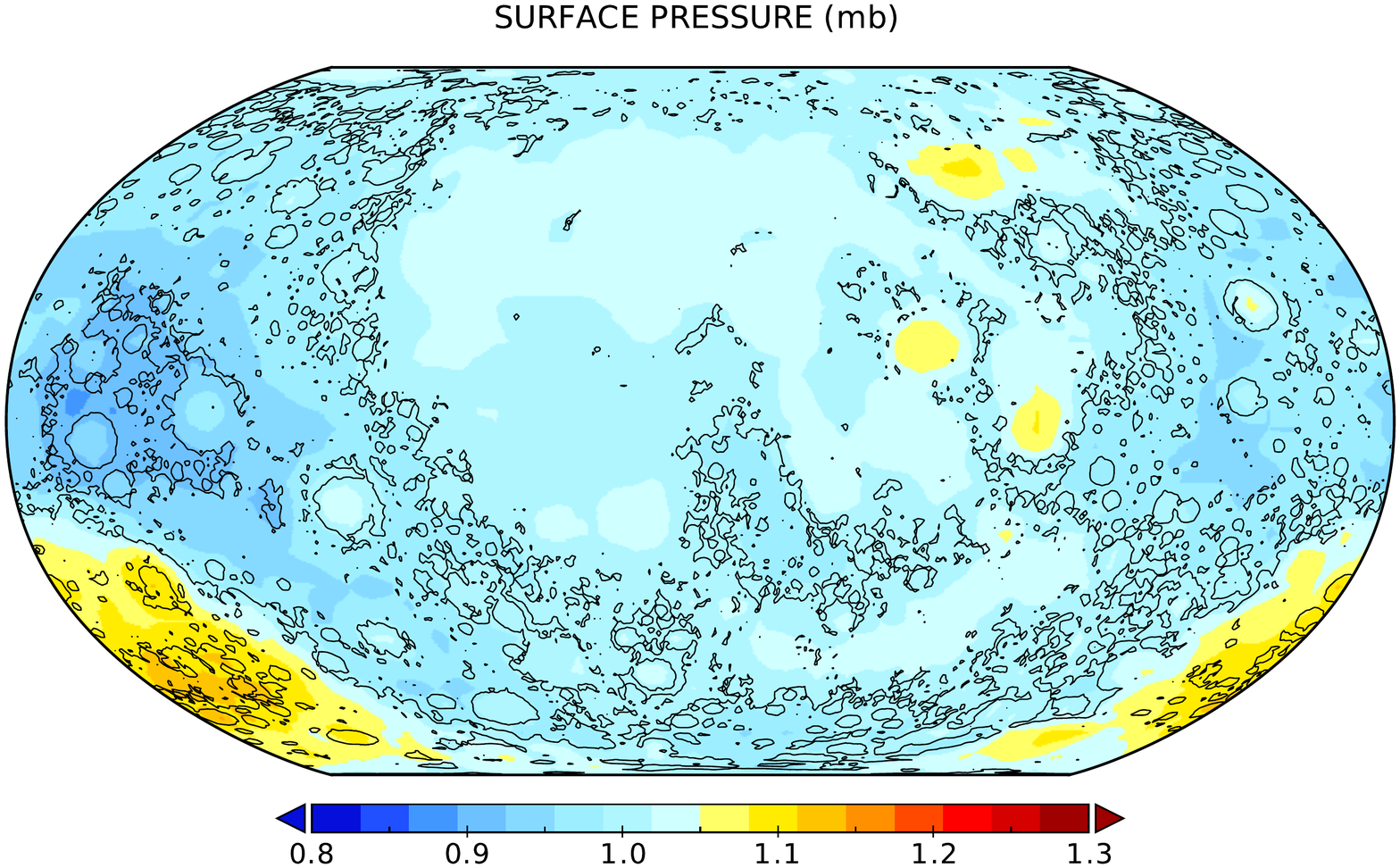}
\includegraphics[trim={0 2cm 0 0},clip,scale=0.28]{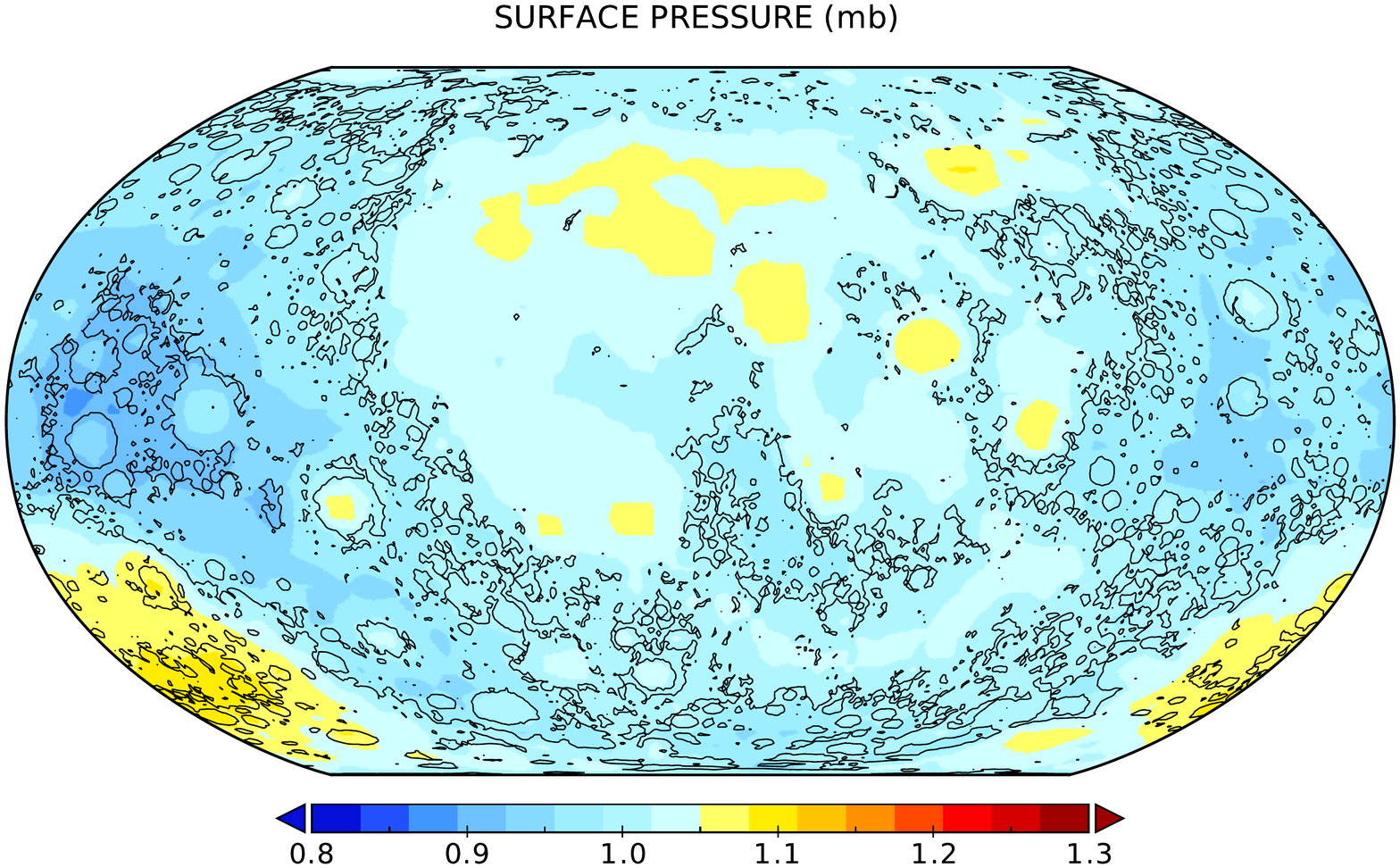}\\

\caption{\small 1 mb ``dry'' CO atmosphere (experiment 1c from Table 2). Instantaneous maps of ground temperature (top row), lower atmospheric layer (2\% mass) temperature (middle row) and surface pressure (bottom row) for the substellar point in the middle of the map (left column) and at the opposite side of the planet (right column)}.
\end{figure*}

\begin{figure*}[ht!]
.\\
\includegraphics[trim={0 2cm 0 0},clip,scale=0.28]{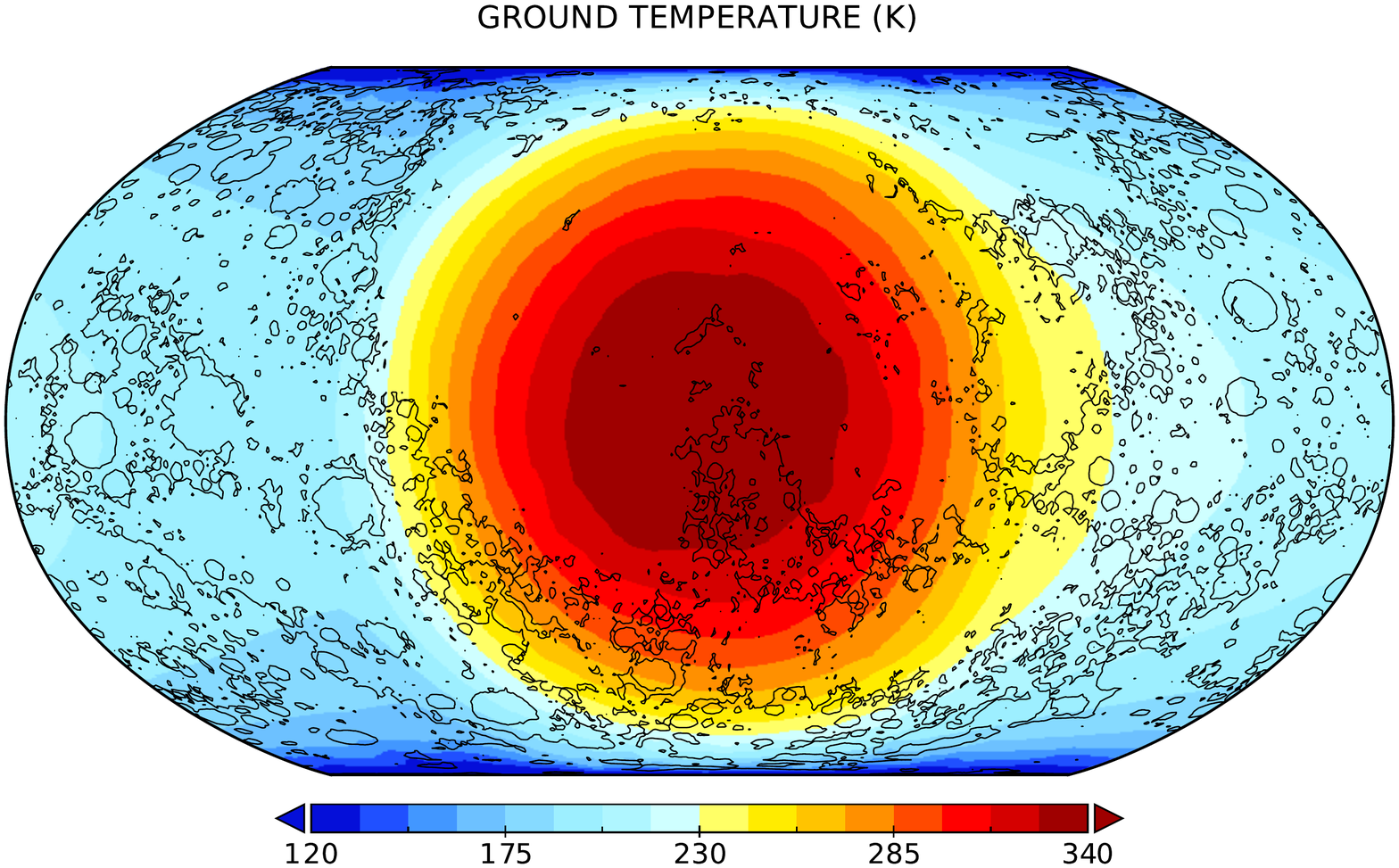}
\includegraphics[trim={0 2cm 0 0},clip,scale=0.28]{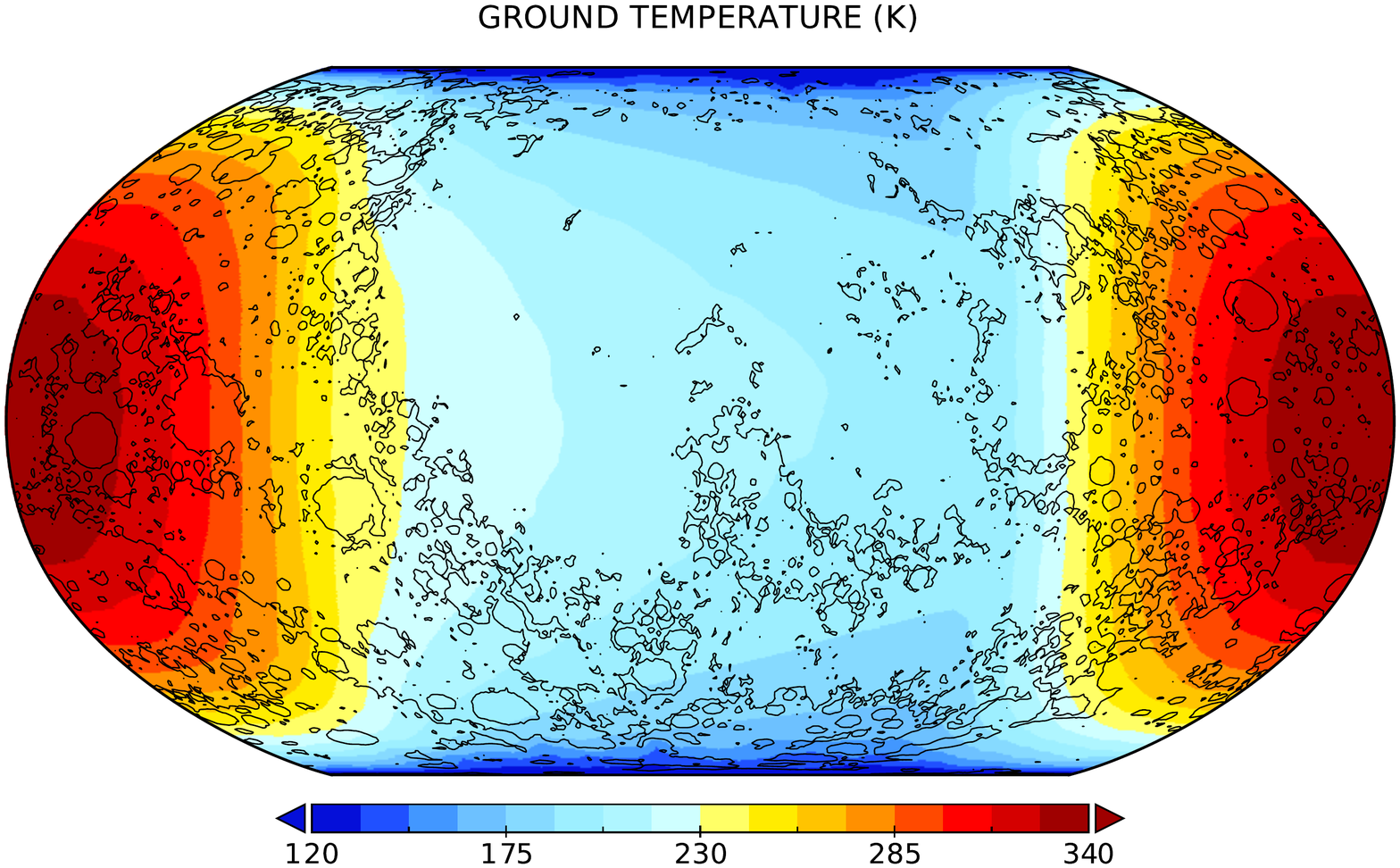}\\
\includegraphics[trim={0 2cm 0 0},clip,scale=0.28]{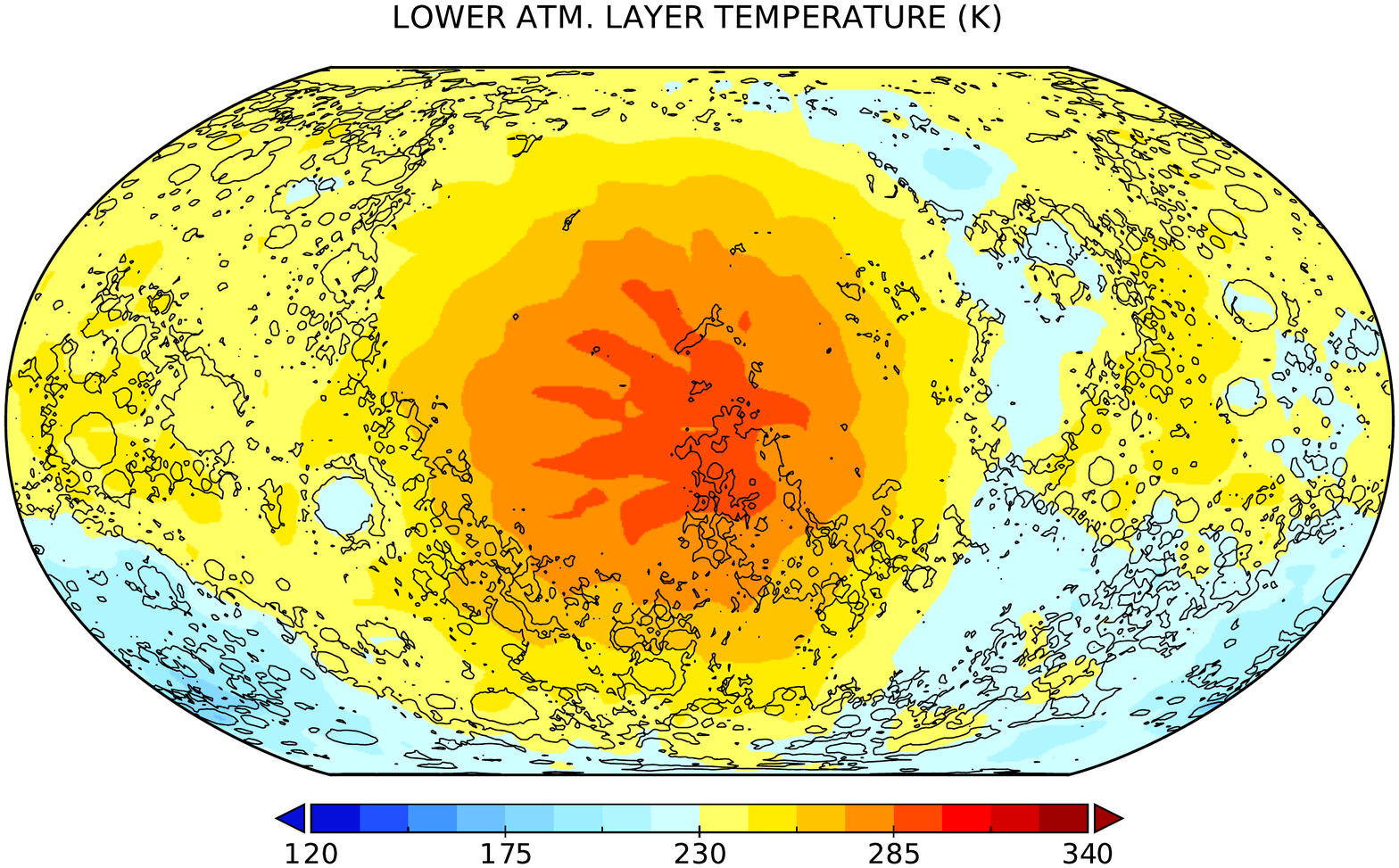}
\includegraphics[trim={0 2cm 0 0},clip,scale=0.28]{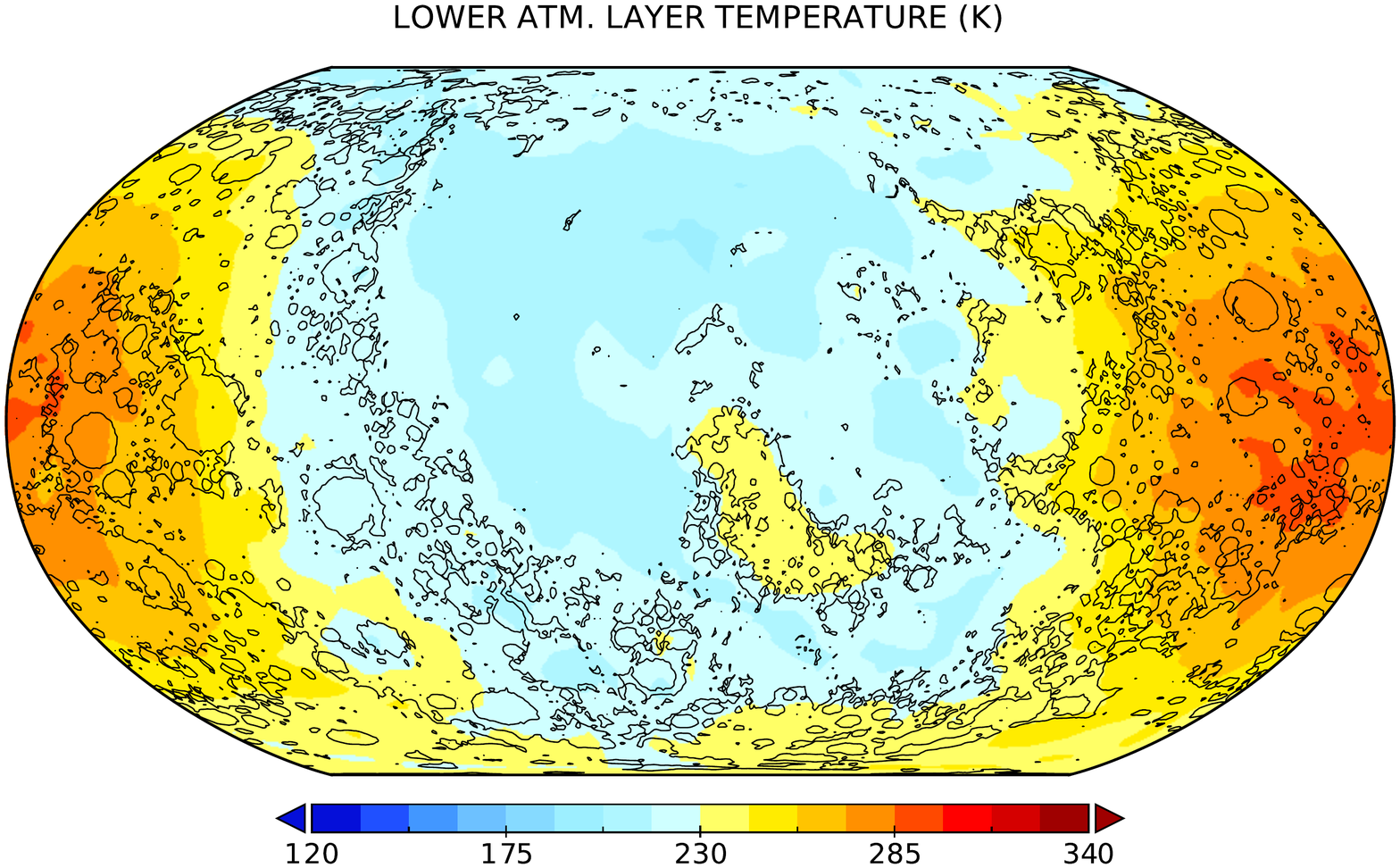}\\
\includegraphics[trim={0 2cm 0 0},clip,scale=0.28]{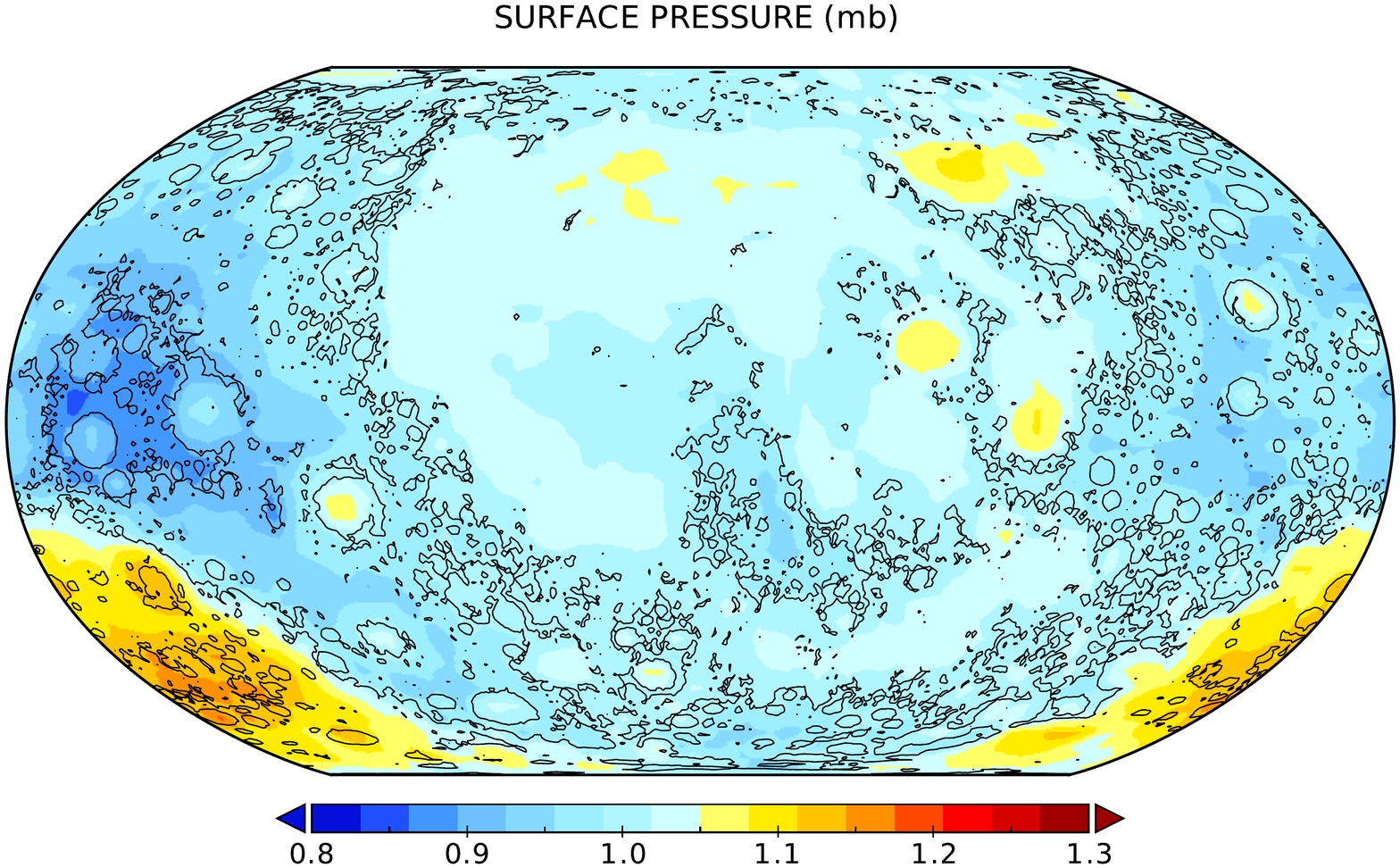}
\includegraphics[trim={0 2cm 0 0},clip,scale=0.28]{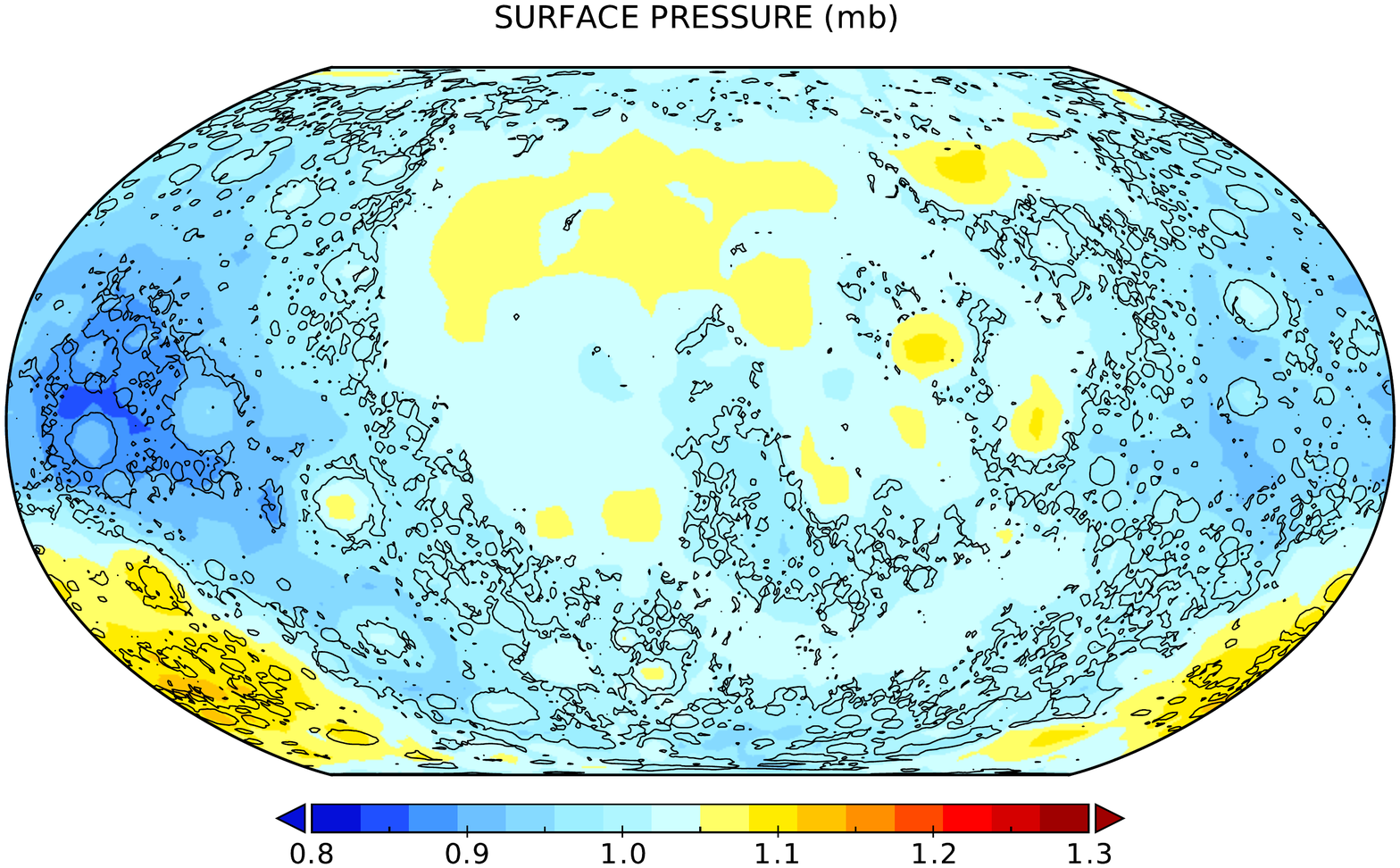}\\

\caption{\small 1 mb ``wet'' CO atmosphere (experiment 3c from Table 2). Instantaneous maps of ground temperature (top row), lower atmospheric layer (2\% mass) temperature (middle row) and surface pressure (bottom row) for the substellar point in the middle of the map (left column) and at the opposite side of the planet (right column)}.
\end{figure*}


\begin{figure*}[ht!]
.\\
\includegraphics[trim={0 2cm 0 0},clip,scale=0.28]{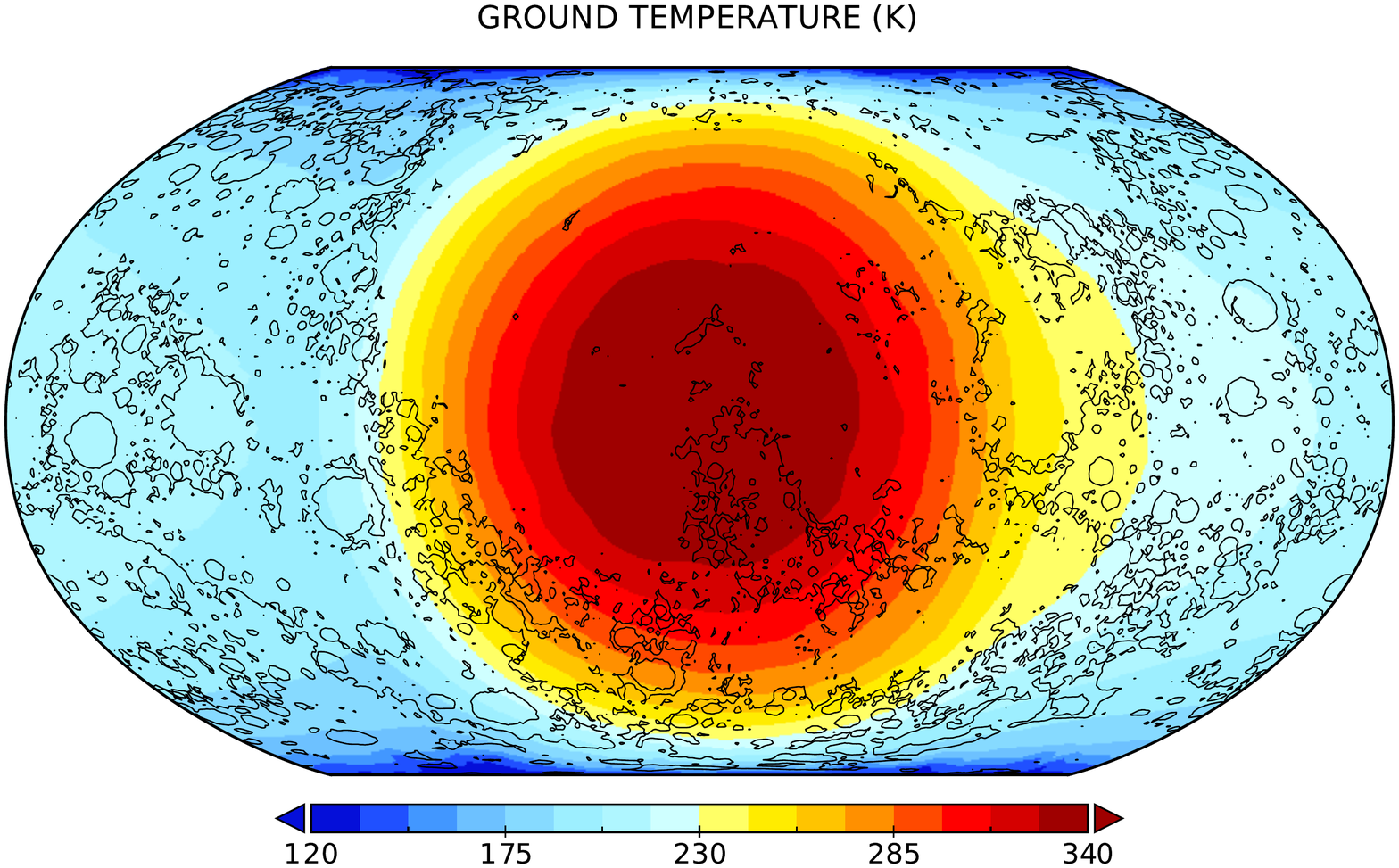}
\includegraphics[trim={0 2cm 0 0},clip,scale=0.28]{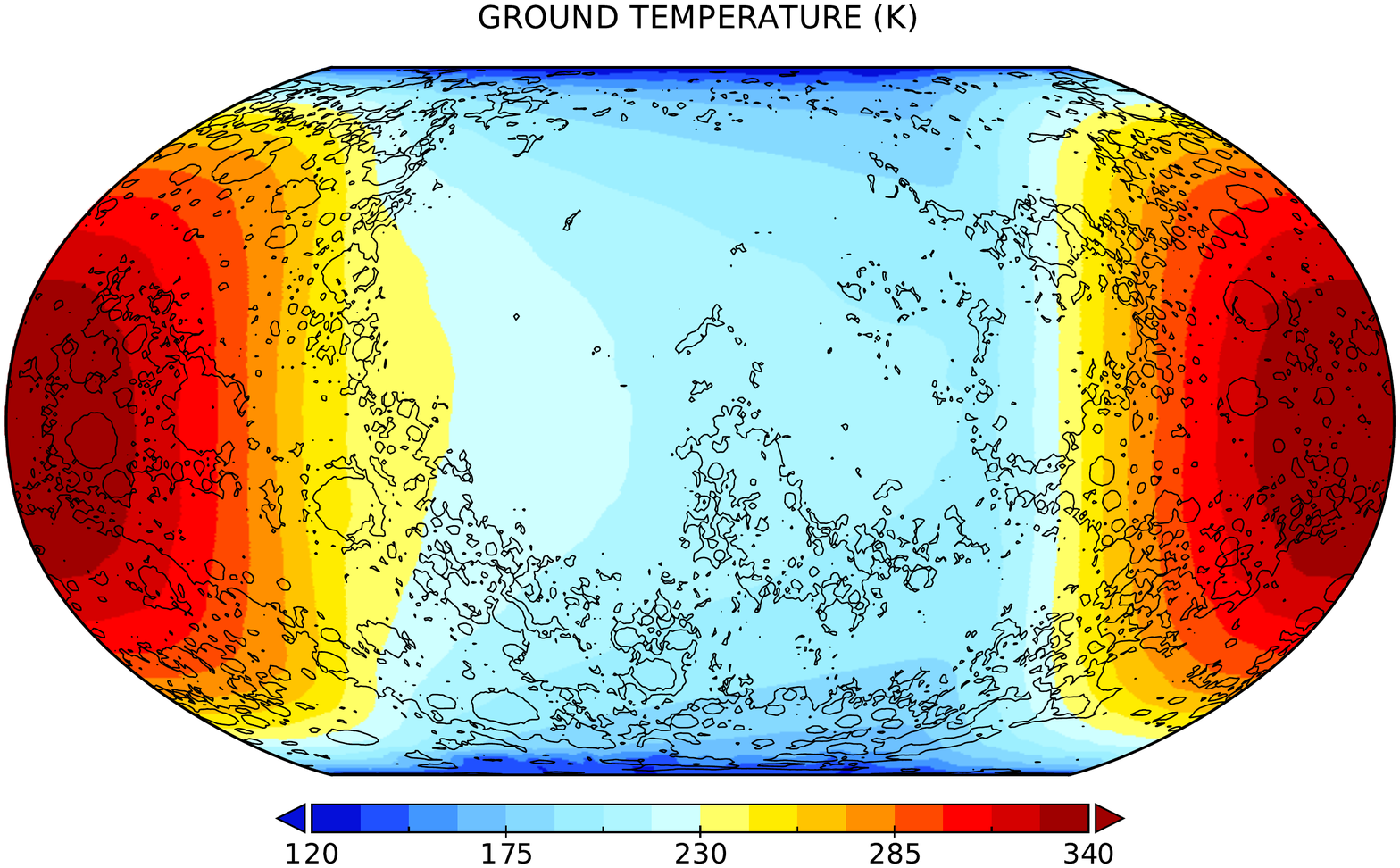}\\
\includegraphics[trim={0 2cm 0 0},clip,scale=0.28]{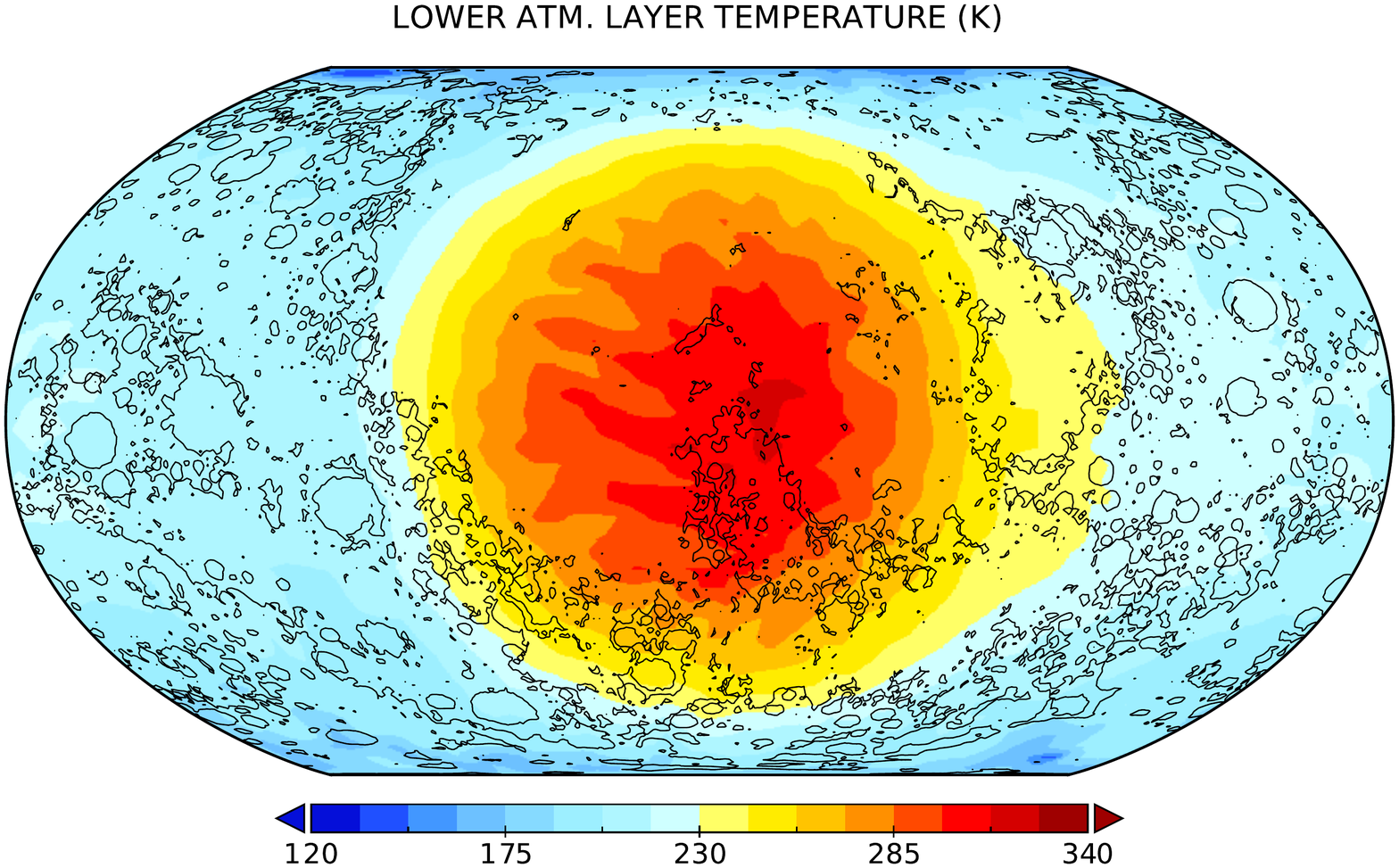}
\includegraphics[trim={0 2cm 0 0},clip,scale=0.28]{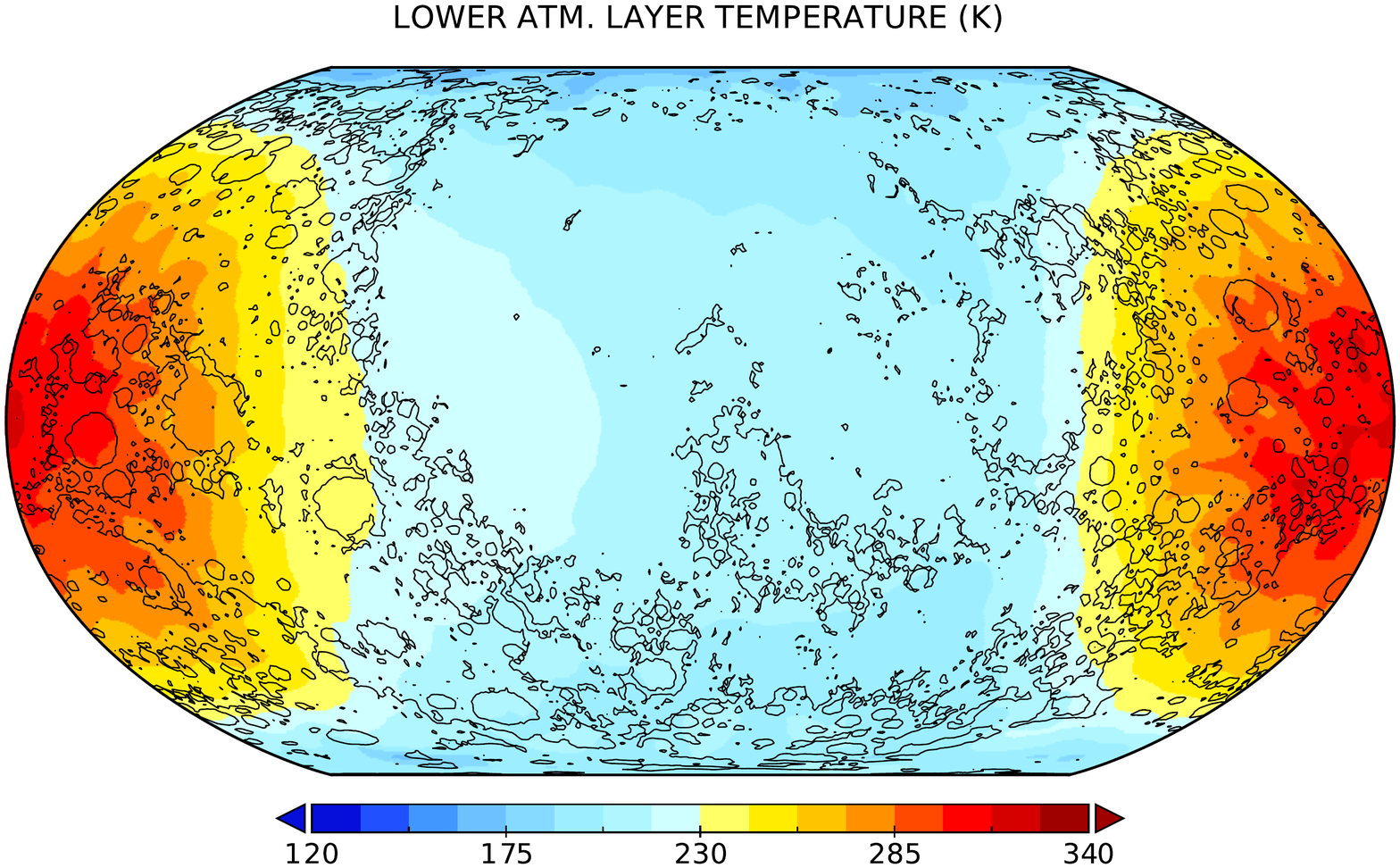}\\
\includegraphics[trim={0 2cm 0 0},clip,scale=0.28]{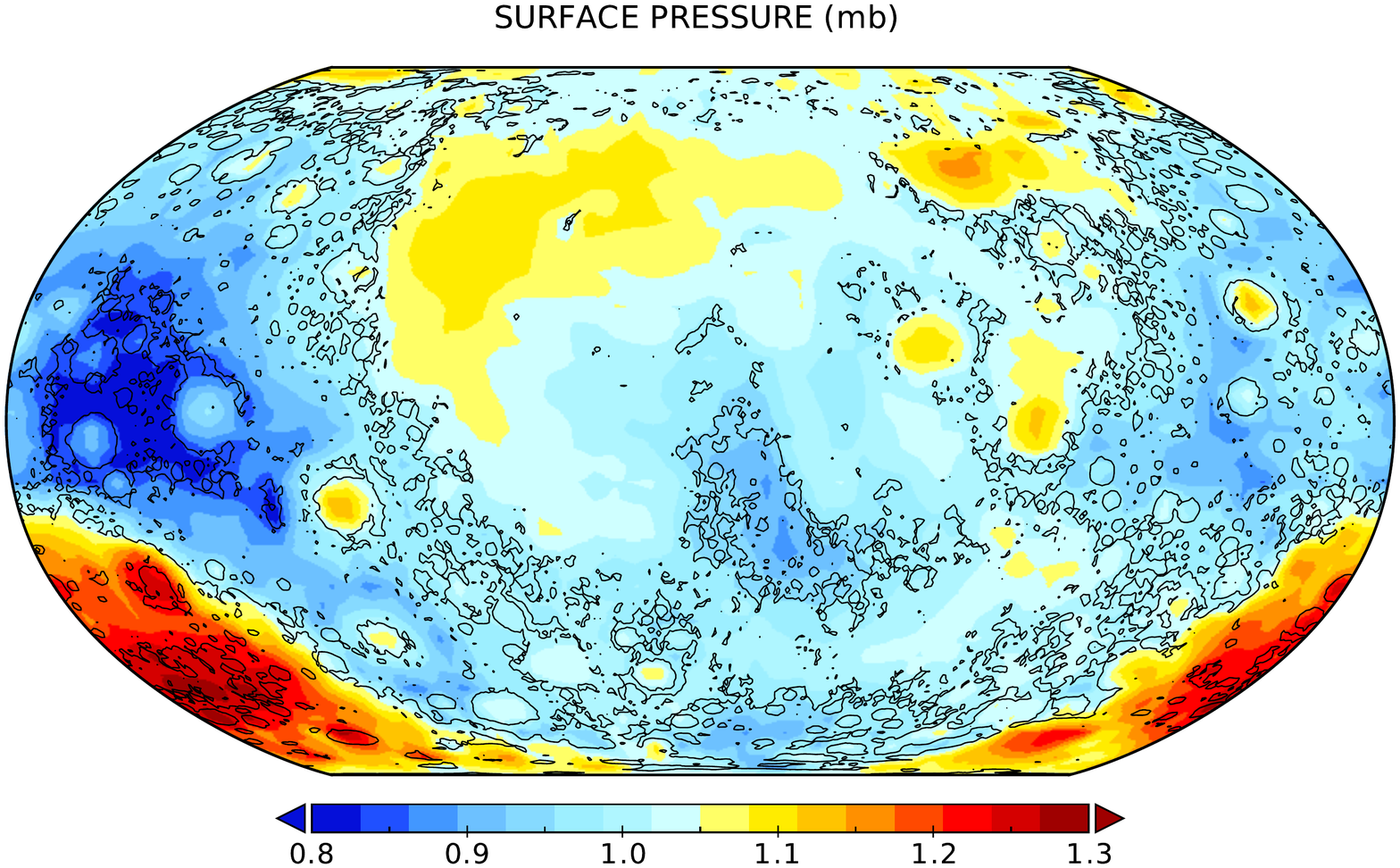}
\includegraphics[trim={0 2cm 0 0},clip,scale=0.28]{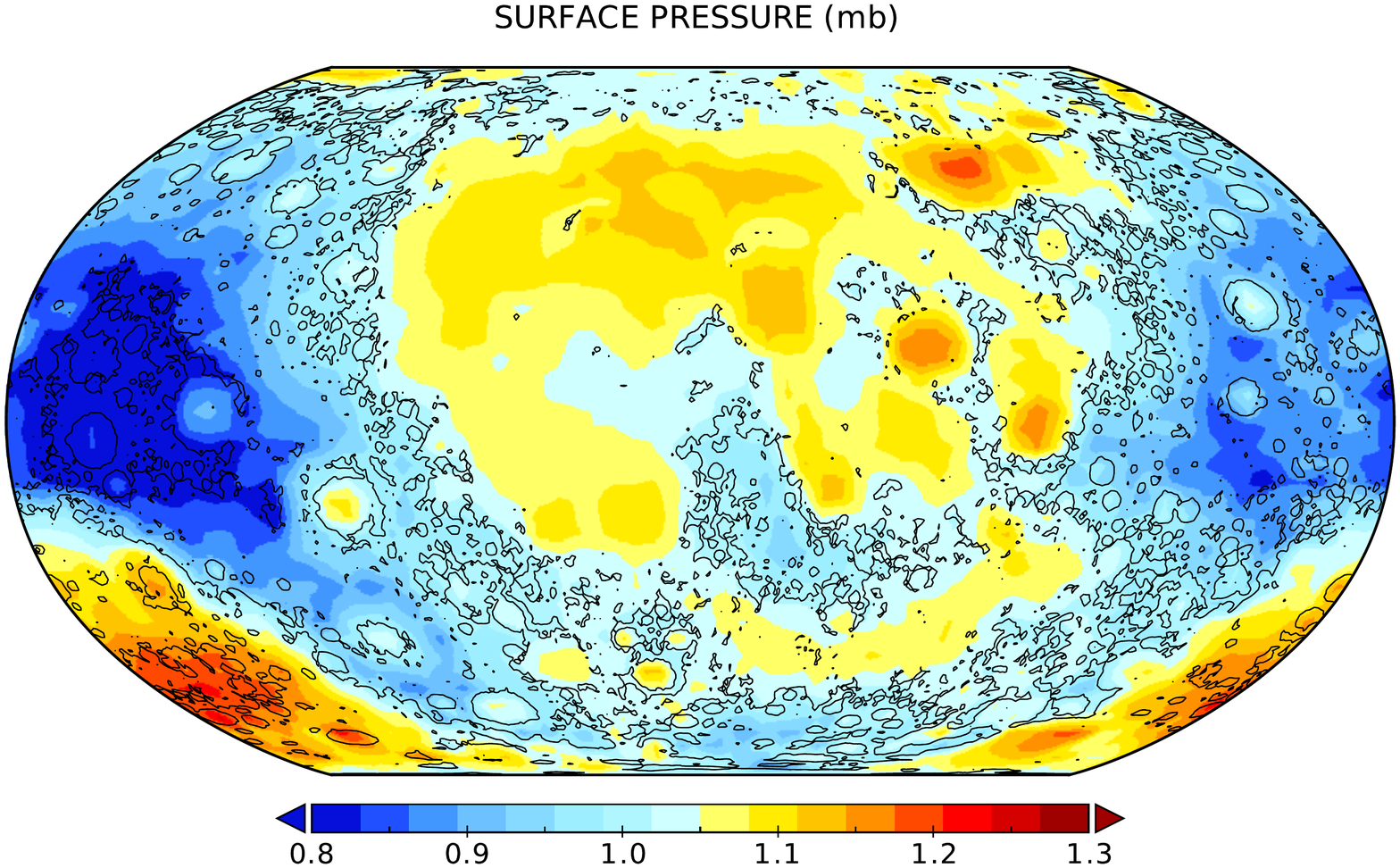}\\

\caption{\small 1 mb ``wet'' CO$_2$ atmosphere (experiment 4c from Table 2). Instantaneous maps of ground temperature (top row), lower atmospheric layer (2\% mass) temperature (middle row) and surface pressure (bottom row) for the substellar point in the middle of the map (left column) and at the opposite side of the planet (right column)}.
\end{figure*}


\end{document}